\documentclass[fleqn,usenatbib]{mnras}

\usepackage{newtxtext,newtxmath}

\usepackage[T1]{fontenc}

\DeclareRobustCommand{\VAN}[3]{#2}
\let\VANthebibliography\thebibliography
\def\thebibliography{\DeclareRobustCommand{\VAN}[3]{##3}\VANthebibliography}

\usepackage{graphicx}	
\usepackage{amsmath}	
\usepackage{array}
\usepackage{booktabs}
\usepackage{threeparttable}
\usepackage{multirow} 
\usepackage{xcolor}
\usepackage{tabularx}
\usepackage{rotating} 
\usepackage{blindtext}
\usepackage{makecell}
\usepackage{orcidlink}
\usepackage{subfigure}
\usepackage{listings}
\usepackage{hyperref}

\newcommand{\Msun}{\ensuremath{\mathrm{M}_\odot}}
\newcommand{\Mstar}{\ensuremath{\mathrm{M}_*}}

\title[BBH Hosts in the 3G Era]{Identifying Host Galaxies of Binary Black Hole Mergers with Next-Generation Gravitational Wave Detector Networks}

\author[S. Biswas et al.]{
Sumedha Biswas$^{1}$\orcidlink{0009-0006-7543-1544}\thanks{E-mail:s.biswas@astro.ru.nl}, Andrew Levan$^{1,2}$, Peter G. Jonker$^{1}$\orcidlink{0000-0001-5679-0695},
Kendall Ackley$^{2}$\orcidlink{000-0002-8648-0767},
Gregory Ashton$^{3}$\orcidlink{0000-0001-7288-2231},
\newauthor{Nikhil Sarin$^{4,5}$\orcidlink{0000-0003-2700-1030}}
\\
$^{1}$ Department of Astrophysics/IMAPP, Radboud University, PO Box 9010, 6500 GL Nijmegen, The Netherlands\\
$^{2}$ Department of Physics, University of Warwick, Coventry, CV4 7AL, UK \\
$^{3}$Department of Physics, Royal Holloway University of London, Egham, TW20 0EX \\
$^{4}$Kavli Institute for Cosmology, University of Cambridge, Madingley Road, CB3 0HA, UK\\
$^{5}$Institute of Astronomy, University of Cambridge, Madingley Road, CB3 0HA, UK\\
}

\date{Accepted XXX. Received YYY; in original form ZZZ}

\pubyear{2025}

\begin{document}
\label{firstpage}
\pagerange{\pageref{firstpage}--\pageref{lastpage}}
\maketitle

\begin{abstract}
Identifying the host galaxy of a binary black hole (BBH) merger detected via gravitational waves (GWs) remains a challenge due to the absence of electromagnetic counterparts and the large localization volumes produced by current-generation detectors. A confident host association would provide stellar population properties to constrain BBH formation channels and enable measurements of cosmological parameters such as the Hubble constant, $H_0$. We simulate BBH mergers in nearby ($z<0.25$) host galaxies to evaluate the feasibility of host identification with future GW detector networks, including configurations with the planned LIGO-India detector and third-generation detectors such as the Einstein Telescope (ET) and Cosmic Explorer (CE). We construct two injection grids to explore variations in BBH mass, distance, and directional sensitivity, and infer localization volumes using the Fisher Information Matrix (FIM)-based parameter estimation implemented through \texttt{BILBY}. To assess the prospects for unique host identification, we introduce a set of diagnostics: theoretical comoving volume thresholds for galaxies of a given stellar mass, derived from galaxy stellar mass functions (GSMFs), a metallicity-based volume threshold motivated by progenitor environment models, stellar mass fractions to quantify candidate host prominence, and the probability of chance alignment ($p_c$). These metrics provide ways to evaluate host associations and constrain BBH formation channels. We find that future networks that include ET and CE localize BBH mergers to volumes smaller than those theoretical thresholds, implying potentially unique host identification, out to $\sim$1000~Mpc at a rate of $\sim 100~{\rm yr^{-1}}$. While associations for individual events may remain uncertain, our framework is well-suited to population-level analyses, enabling constraints on BBH formation scenarios in the era of next-generation GW detector networks.
\end{abstract}



\section{Introduction}
The discovery of gravitational waves (GW) from a binary black hole (BBH) merger was made during the first observing run (O1) of the two LIGO detectors at Hanford and Livingston (USA) in 2015, marking the beginning of GW astrophysics \citep{gw150914}. Since then, the two LIGO detectors have undergone further upgrades in sensitivity \citep{2018abbott, 2025capote, gwtc4_intro}, and Advanced Virgo \citep{virgo2015, virgo_improve1, virgo_improve2} in Italy and KAGRA \citep{kagra_2013, kagra2019, kagra_2020} in Japan have been operational, since 2017 and 2020, respectively. With this combined GW detector network, many BBH mergers and other compact binary coalescences (CBCs) have been detected across successive observing runs. Combining data from the GW transient catalogues (GWTC-1.0: \citealt{gwtc1}, GWTC-2.0, 2.1: \citealt{gwtc_2a, gwtc_2b}, GWTC-3.0: \citealt{gwtc3}, and GWTC-4.0: \citealt{gwtc4}), there are 218\footnote{ GW candidates with a probability of astrophysical origin $\rm p_{astro} \geq 0.5$;} GW event candidates in total, comprising 210 BBH mergers, 2 binary neutron star (BNS) mergers, and 6 neutron star–black hole (NSBH) mergers. 



GW transients, involving at least one neutron star (NS), i.e. BNS and NSBH mergers, can produce a range of electromagnetic (EM) counterparts such as gamma-ray bursts \citep[e.g.][]{eichler89}, kilonovae \citep[e.g.][]{li98, Metzger_2012, tanvir2013, 2017Coulter, Smartt_2017, 2022rastinejad, troja2022, 2024levan}, and potentially fast X-ray transients \citep[e.g.][]{siegelciolfi_1, siegelciolfi_2, sun2017, sun2019, sun2023magnetar, quirola_bns, biswas2025}. These signatures are thought to arise from the accretion of a remnant disk onto the massive NS or BH remnant following the merger \citep[e.g.][]{1992narayan}, radioactive decay of heavy, neutron-rich, elements synthesised through the $r$-process in the expanding merger ejecta \citep{optical_counterparts_1998, Tanvir_2017}, or the spin-down energy of a milli-second magnetar (e.g. \citealt{Lin_2022}). In the case of NSBH mergers, the outcome depends on the binary parameters \citep[e.g.][]{2005davies, 2013foucart}: the NS may be tidally disrupted outside the event horizon of the BH, leading to substantial mass ejection, or it may plunge directly into the BH with little or no ejecta. Since their GW signals are typically louder than those of BNS mergers, and tidal disruption can eject more material, giving rise to brighter kilonovae, NSBH mergers are regarded as especially promising targets for EM follow-up \citep{2023gupta}. During O3, two NSBH mergers \citep{lvk_nsbh} were detected; however, neither yielded an EM counterpart. In contrast, BBH mergers, are generally not expected to produce EM emission, although possible EM bright scenarios do exist, for example if the merger occurs in a gas-rich environment such as the accretion disks of active galactic nuclei (AGN) \citep[e.g.][]{Perna_2016, McKernan_2019, agn_bbh_det}. Notably, the first detected BBH merger, GW150914 \citep{gw150914}, was temporally and spatially coincident with a weak gamma-ray burst detected by \textit{Fermi}-GBM \citep{gw150914_fermi, gw150914_fermigbm} although no optical or radio emission was detected \citep{Savchenko_2016}, and this association remains debated \citep[e.g.][]{agnieszka2017, conn2018}. In another BBH merger, GW190521 \citep{gw190521}, a contemporaneous optical flare lasting several months from an active galactic nucleus within the GW localization volume was found, with \citet{agn_bbh_det} suggesting the possibility of it occurring inside the accretion disk of a central supermassive BH. However, these claims remain controversial, and are not generally accepted at the present time \citep[e.g.][]{Palmese_2021, 2025niccolo}. Hence, at present there are no clear EM counterparts, or indeed precise localizations and host galaxy associations, for any BBH mergers. 

One of the primary challenges in GW astronomy is the precise localization of CBCs, as GW skymaps often span up to hundreds or thousands of square degrees on the sky \citep{Gehrels_2016, Abbott_2020_locali, 2020pankow}. The identification of an EM counterpart significantly improves the localization, as demonstrated in the case of GW170817 \citep{Abbott_2017}, a BNS merger that occurred at a nearby distance of 40~Mpc and was associated with a range of EM counterparts across the spectrum \citep{2017Andreoni, 2017Arcavi, 2017Chornock, 2017Coulter, 2017Covino, 2017Cow, 2017Drout, 2017Kasliwal, 2017Evans, 2017Lipunov, 2017Nicholl, 2017Pian, 2017Shappee, 2017Smartt, 2017Tanvir, 2017Troja, Yang_2019, 2017utsumi, Soares-Santos_2017}, and was subsequently successfully associated with its host, NGC~4993 \citep{2017levan, Im_2017, 2020ebrov, Kilpatrick_2022}. Conversely, in the case of BBH mergers, which are generally assumed to have no associated EM counterparts, GW localization and host identification become extremely challenging since they must be undertaken purely on the basis of GW information. However, it is astrophysically important. The properties of the BBH environment play a critical role in constraining both formation and merger pathways (see \citealt{Mapelli2020} for a detailed review). For instance, low-metallicity environments may favour the formation of more massive stellar-origin BBHs \citep[e.g.][]{2016lamberts}, while dense stellar regions such as globular clusters may support hierarchical BH assembly through repeated dynamical interactions \citep[e.g.][]{2019antonini, 2020martinez, 2022ye}. A uniquely identified host galaxy (or even location within the host galaxy) provides direct measurements of the underlying stellar population properties such as total stellar mass, age of the dominant population and metallicity that can be directly compared to the predictions of different formation channels. Beyond constraining these channels, host identification also has cosmological applications; for instance, it provides a method for an independent redshift measurement to be combined with the luminosity distance inferred from GW observations, enabling a direct measurement of the Hubble constant \citep[e.g.][]{2025gwtc4_hubbleconstant}. Since BBH mergers can be found at much larger distances than those containing NSs, they probe well into the Hubble flow and can, therefore, enable better constraints on expansion histories with fewer observations by, for example, minimising uncertainties due to peculiar motions that were substantial in the case of GW170817 \citep{2017hjorth, 2018cantiello}. In the absence of host identification, several studies have utilized BBH mergers as statistical dark sirens to place constraints on the Hubble constant \citep[e.g.][]{2008macleod, 2012delpozzo, 2022gray, 2024bom, 2024ghosh}, but even a small number of unique host identifications would provide great diagnostic power. 

Although localizing BBH mergers remains challenging with current detector sensitivities, it may be possible in some circumstances \citep{nuttall10,chen16,howell18}, and this limitation may not persist in the coming decades. By the mid-2030s, a range of developments is expected to dramatically improve GW sky localization. Future detections of stellar-mass BBH mergers may be preceded by months to years of inspiral signals in the millihertz regime \citep{2021ewing}, detectable by the space-based Laser Interferometer Space Antenna (LISA) detector \citep{2017amaro, 2024lisa}, allowing advanced pre-merger localization. On the ground, localization prospects will improve with the planned addition of LIGO-India \citep{indigo1, indigo2, 2024ligoa} to the existing network of advanced detectors. Crucially, third-generation (3G) detectors such as the Einstein Telescope (ET) \citep{Punturo_2010, 2011hild, 2020maggiore, etbluebook} and Cosmic Explorer (CE) \citep{2019reitze, 2021evans, 2023evans} are predicted to have an order-of-magnitude improvement in sensitivity \citep{2024bor}. When operating in conjunction with 2G detectors, these facilities will potentially enable precise 3D localization for nearby, high signal-to-noise ratio (SNR) BBH events, achieving sky areas $\rm < 1~deg^2$ \citep{2024bor, Gupta_2024, 2024mo}.

In this paper, we investigate the feasibility of identifying hosts of BBH mergers using future GW detector networks. We simulate BBH mergers injected into \Mstar~galaxies and construct two complementary injection grids, Grids I and II, to explore how localization performance depends on source properties (mass and distance) and directional sensitivity (Section~\ref{inj_grids}). We first define three GW detector network configurations: HLV, HLVKIEC, and EC (Section~\ref{gwdetnets}); and then perform parameter estimation using the Fisher Information Matrix (FIM) formalism implemented in \texttt{BILBY} (Section~\ref{pe_section}). The resulting posteriors are used to compute 3D localization volumes for each injection (Section~\ref{localizationvolume}). We then compare these volumes to theoretical comoving volume thresholds derived from galaxy stellar mass functions, including versions weighted by assumptions about isolated and dynamical BBH formation scenarios (Section~\ref{gsmf_mod}). To further assess the astrophysical utility of these volumes, we introduce a set of diagnostics: a metallicity-based threshold (Section~\ref{metallicity}), mass fractions (Section~\ref{massratios}), and probability of chance alignment (Section~\ref{pchance}). In Section~\ref{results}, we present detailed analyses of localization performance across both injection grids, using each of the proposed diagnostics. These analyses are used to compare detector network performance (Section~\ref{networkcomparison}) and to evaluate the extent to which hosts may be identified and BBH formation channels statistically constrained (Section~\ref{implications}). We summarize our findings and discuss future directions in Section~\ref{conclusion}. In Appendix~\ref{gwparams}, we explore how our assumptions about GW source parameters influence the analysis outcomes and compare our adopted values to those inferred from observed GW events. 

\section{Methods} \label{methods}
We adopt a flat $\Lambda$CDM cosmology with a matter density parameter $\Omega_m = 0.3$ throughout this analysis. For sections of the study requiring observed galaxy information, we use the NED Local Volume Sample (NED-LVS; \citealt{2023cook}), a subset of $\sim$2 million objects in NED with distances out to 1000~Mpc. For bright galaxies ($\geq L^*$), NED-LVS is $\sim$100\% complete out to $\sim$400~Mpc \citep{2023cook}. The catalogue adopts $H_0 = 69.6~\rm km~s^{-1}~Mpc^{-1}$, which we also use in relevant calculations to ensure consistency.

\subsection{GW Detector Networks} \label{gwdetnets}
We consider the following GW detectors in our analysis: \textit{current-generation detectors}— LIGO-Hanford (H), LIGO-Livingston (L) \citep{Abbott_2009}, Virgo (V) \citep{virgo2015}, and KAGRA (K) \citep{kagra2019}; and \textit{proposed third-generation detectors}— Einstein Telescope (E) \citep{Punturo_2010, 2011hild, 2020maggiore, etbluebook}, Cosmic Explorer (C) \citep{2019reitze, 2021evans, 2023evans}, and LIGO-India (I) \citep{indigo1, indigo2, 2024ligoa}. The specifications of each detector, such as the location, frequency range, and orientation, are listed in Table~\ref{tab:detectors}. Based on these, we define three GW detector networks that we use throughout this study: HLVKIEC, HLV, and EC. For each detector, we use publicly available power spectral density (PSD) data\footnote{Design Sensitivity curves have been downloaded from: \hyperlink{https://dcc.ligo.org/ligo-t2000012/public}{LIGO-T2000012-v2}, \hyperlink{https://dcc.cosmicexplorer.org/CE-T2000017/public}{\href{https://dcc.cosmicexplorer.org/CE-T2000017/public}{CE-T2000017-v8}}, and \hyperlink{https://www.et-gw.eu/index.php/etsensitivities}{ET}}. 

\begin{figure*}
    \centering
    \includegraphics[width=\linewidth]{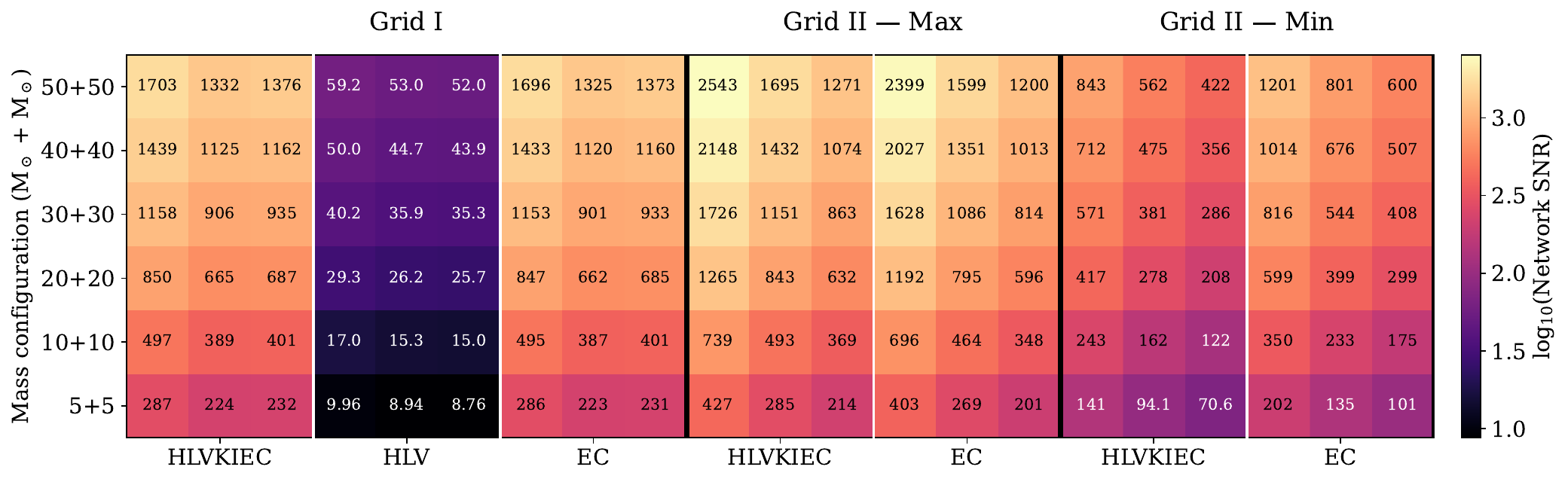}
    \caption{Heatmap of the network optimal SNRs for the injected BBH mergers across masses and distances. \textit{Black} vertical lines separate Grids I and II (maximum and minimum sensitivity), and the \textit{white} vertical lines separate the GW detector networks. Each triplet of columns within a network corresponds to distances of 500, 750, and 1000 Mpc \textit{(left to right)}, and rows correspond to mass configurations from 50+50~\Msun~\textit{(top)} to 5+5~\Msun~\textit{(bottom)}. The colourbar indicates the logarithm of the network optimal SNR, with lighter shades corresponding to higher SNR values. The network optimal SNR is calculated as $\rho_{\rm network} = \sqrt{\sum_j \rho_j^2}$, where $\rho_j$ is the optimal matched-filter SNR in detector $j$.}
    \label{fig:networksnrs}
\end{figure*}

\begin{table*}
\centering
\caption{GW detector network specifications.}\label{tab:detectors}
\begin{tabular}{lllll}
\toprule \hline
\textbf{Detector} & \textbf{Location (Lat, Lon)} & \textbf{Frequency Range (Hz)} & \textbf{Arms} & \textbf{Orientation (Azimuths)} \\
\midrule
LIGO Hanford (H) & 46.45°N, 119.41°W & 20--2000 & 2 & 36.8°, 126.8° \\
LIGO Livingston (L) & 30.56°N, 90.77°W & 20--2000 & 2 & 108.0°, 198.0° \\
Virgo (V) & 43.63°N, 10.50°E & 20--2000 & 2 & 19.4°, 109.4° \\
KAGRA (K) & 36.41°N, 137.30°E & 20--2000 & 2 & 25°, 295° \\
LIGO-India (I) & 19.61°N, 77.02°E & 20--2000 & 2 & TBD \\
Einstein Telescope (ET) & 50.85°N, 5.70°E & 1--10,000 & 3 & 0°, 120°, 240° \\
Cosmic Explorer (CE) & 46.45°N/30.56°N, & 5--5,000 & 2 & Site-dependent \\
& 119.41°W/90.77°W & & & \\
\bottomrule
\end{tabular}
\vspace{0.2cm}
\parbox{\textwidth}{
\footnotesize
\textbf{Notes:} Orientation angles are azimuths measured clockwise from geographic north. L-shaped detectors have 90° arm separations. ET is planned to have three 10 km arms arranged in an equilateral triangle. CE's final orientation depends on site selection (40 km arms). Frequency ranges represent approximate design goals.
}
\end{table*}

\subsection{Injection Grids} \label{inj_grids}
With the detector networks established, we proceed to construct two complementary grids of GW injections to evaluate how both sky location and detector sensitivity affect localization performance.

\begin{figure}
    \centering
    \includegraphics[width=\linewidth]{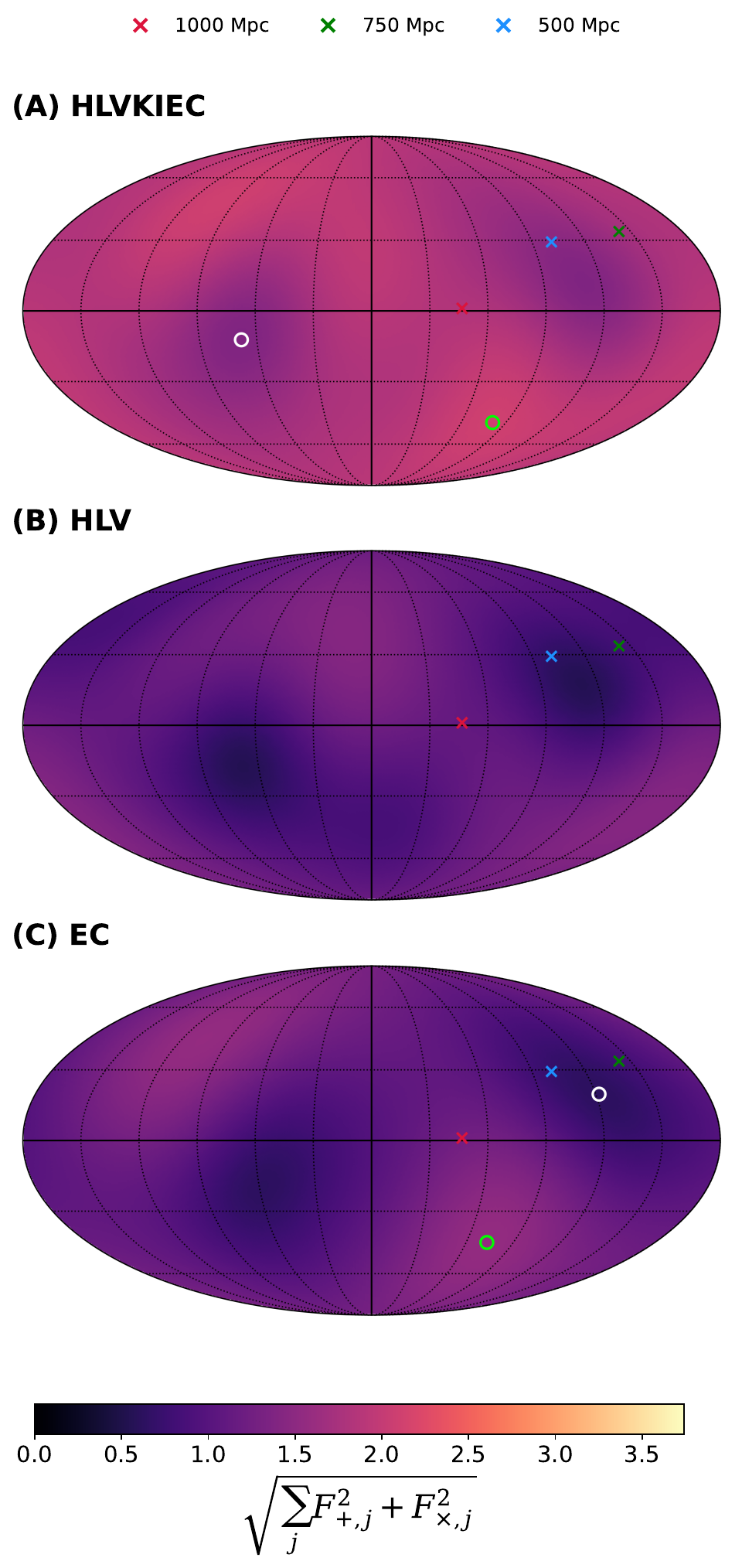}
    \caption{Antennae pattern maps for the three GW detector networks considered in this study: (A) HLVKIEC, (B) HLV, and (C) EC. Each panel shows the combined network antenna amplitude response, defined as $\sqrt{\sum_j F_{+,j}^2 + F_{\times,j}^2}$ \citep{finn2001, 2011schutz}, as a function of sky position in equatorial coordinates at time t. The red, green, and blue crosses mark the locations of the three injected \Mstar~galaxies at 1000~Mpc, 750~Mpc, and 500~Mpc (Table~\ref{tab:injected_hosts}), respectively, used in Grid I. The lime green and white circles denote the “bright” (maximum sensitivity) and “dark” (minimum sensitivity) sky locations in panels (A) and (C) (Table~\ref{tab:bright_dark_points}), respectively, selected based on the network antenna pattern and used in Grid II.}
    \label{fig:antennapattern}
\end{figure}

\begin{enumerate}
    \item \textbf{Grid I:} We select 3 \Mstar~galaxies—systems with typical total stellar masses near the characteristic “knee” of the double Schechter galaxy stellar mass function (Equation~\ref{schechter}), marking the transition from the abundant low-mass population to the rarer massive galaxies—from the NED-LVS galaxy catalogue, located at distances of $\sim$500~Mpc, $\sim$750~Mpc, and $\sim$1000~Mpc respectively (Table~\ref{tab:injected_hosts}). These distances are similar to several of the BBH detections reported in GWTC 4.0 \citep{gwtc1, gwtc_2a, gwtc_2b, gwtc3, gwtc4} and are sufficiently close that well-constrained skymaps maintain reasonably small comoving volume regions that can be realistically surveyed for galaxies with existing telescopes. In each galaxy, we simulate a BBH merger with one of the following component mass configurations: 50+50~\Msun, 40+40~\Msun, 30+30~\Msun, 20+20~\Msun, 10+10~\Msun, and 5+5~\Msun. This results in an injection grid with a total dimension of $3 \times 3 \times 6$ for 3 GW detector networks. 

    \item \textbf{Grid II:} To explore the “best” and “worst” case scenarios for detector network performance and compare the performance of the HLVKIEC and EC detector networks, we incorporate directional sensitivity using the network antenna patterns. The response of a GW detector to a signal depends on its antenna pattern functions $F_{+}$ and $F_{\times}$, which quantify the sensitivity of the detector to the two independent polarization states of a GW signal \citep{finn2001, 2011schutz}. To construct this grid, we select arbitrary sky locations within the most sensitive ("bright") and least sensitive ("dark") regions of each detector network’s (HLVKIEC and EC) antenna pattern (Figure~\ref{fig:antennapattern},  Table~\ref{tab:bright_dark_points}). At each of these positions, we simulate BBH mergers at the same three distances and BBH mass configurations as in Grid I. This results in an injection grid with a total dimension of $2 \times 2 \times 3 \times 6$ for 2 GW detector networks. We exclude the HLV network from Grid II as the FIM-based parameter estimation (PE; discussed in Section~\ref{pe_section}) becomes unreliable due to low SNRs.
\end{enumerate}

\begin{table*}
    \centering
    \caption{Properties of the injected \Mstar~hosts in Grid I, selected from the NED-LVS galaxy catalogue. For each injection distance, the corresponding characteristic stellar mass \Mstar was determined using the corresponding Schechter function parameters (see Table~\ref{tab:galaxy_mass_function}).}
    \begin{tabular}{lcccc}
    \toprule \hline
    \textbf{Galaxy Name} & \textbf{RA, Dec [deg]} & \textbf{Distance [Mpc]} & \textbf{Redshift} & \textbf{Stellar Mass [\Msun]} \\
    \midrule
    WISEA J171543.62+291712.7 & 258.93177, 29.28687 & 499.922 $\pm$ 0.145 & 0.12 & $6.16 \times 10^{10}$ \Msun \\
    WISEA J142633.23+335512.5 & 216.63848, 33.92015 & 750.081 $\pm$ 0.053 & 0.18 & $6.15 \times 10^{10}$ \Msun \\
    WISEA J205335.01+010024.3 & 313.39591, 1.00678 & 999.938 $\pm$ 0.326 & 0.23 & $4.93 \times 10^{10}$ \Msun \\
    \bottomrule
    \end{tabular}
    \label{tab:injected_hosts}
\end{table*}

\begin{table}
\centering
\caption{Sky coordinates for Grid II, corresponding to the most and least sensitive points on the antenna response pattern for each detector network (HLVKIEC, HLV, and EC). The sensitivity is quantified by the combined antenna power $\sum_j \sqrt{(F_{+,j}^2 + F_{\times,j}^2)}$, computed over the full sky at a fixed geocentric time $t_c = 0.0$ and polarization angle $\psi = 0$. The \textit{maximum} sensitivity represent directions of peak network response to GWs, while the \textit{minimum} sensitivity correspond to minima in directional sensitivity.}
\begin{tabular}{lccccc}
\toprule
\textbf{Network} & \textbf{Sensitivity} & \textbf{RA (deg)} & \textbf{Dec (deg)} \\
\midrule
\multirow{2}{*}{HLVKIEC} 
    & Maximum & 279.000 & $-48.923$ \\
    & Minimum   &  68.203 & $-12.025$ \\
\midrule
\multirow{2}{*}{EC} 
    & Maximum & 286.967 & $-44.202$ \\
    & Minimum   & 238.359 & 19.471 \\
\bottomrule
\end{tabular}
\label{tab:bright_dark_points}
\end{table}


\subsection{Parameter Estimation with \texttt{BILBY}} \label{pe_section}
To perform PE on our simulated GW injection grid, we use the Bayesian inference library \texttt{BILBY} \citep{bilby}. All analyses are performed using the \texttt{IMRPhenomXPHM} waveform approximant \citep{Pratten_2021}, a frequency-domain, precessing, phenomenological model well-suited to modeling the inspiral, merger, and ringdown of non-eccentric BBHs. For each injection, we use Grid I or Grid II values (Section~\ref{inj_grids}, Tables~\ref{tab:injected_hosts}, \ref{tab:bright_dark_points}), while keeping all other parameters fixed across runs (Table~\ref{tab:injection_parameters}). PE is performed using standard astrophysical priors (Table~\ref{tab:prior_ranges}): we assume uniform priors on Right Ascension (RA), polarization angle $\psi$, and coalescence phase $\phi_c$; sinusoidal priors on the inclination angle $\theta_{\rm JN}$; and a cosine prior on Declination (Dec). The luminosity distance prior $d_L$ between 1 and 10000 Mpc, is uniform in comoving volume and source frame time i.e. $\propto \frac{1}{1+z} \frac{dV_c}{dz}$ where $V_c$ is the comoving volume. The coalescence time is allowed to vary within a $\pm$0.5~s window around the injection time. All other parameters, including masses and spins, are fixed to their injected values.

\begin{table}
\centering
\caption{Fixed injection parameters in the \texttt{BILBY} runs. For each run, $m_1 = m_2 = [50.0, 40.0, 30.0, 20.0, 10.0, 5.0] M_\odot$, and the sky position coordinates (RA, Dec) vary according to Tables~\ref{tab:injected_hosts}, \ref{tab:bright_dark_points}.}
\begin{tabular}{lcc}
\toprule \hline
\textbf{Parameter} & \textbf{Injected Value} & \textbf{Units} \\
\hline
Spin magnitude 1, $a_1$            & 0.1           & --        \\
Spin magnitude 2, $a_2$            & 0.1           & --        \\
Tilt angle 1, $\theta_1$           & 0.0           & rad       \\
Tilt angle 2, $\theta_2$           & 0.0           & rad       \\
Azimuthal spin angle, $\phi_{12}$ & 0.0           & rad       \\
Precession angle, $\phi_{\rm JL}$     & 0.0           & rad       \\
Inclination angle, $\theta_{\rm JN}$   & 0.4           & rad       \\
Polarization angle, $\psi$         & 0.0           & rad       \\
Coalescence phase, $\phi_c$        & 0.0           & rad       \\
Geocentric time, $t_c$             & 0.0           & s         \\
\hline
\end{tabular}
\label{tab:injection_parameters}
\end{table}

\begin{table}
\centering
\caption{Priors and their ranges used in the \texttt{BILBY} runs. Parameters not listed here are fixed to their injected values (Table~\ref{tab:injection_parameters}).}
\begin{tabular}{lcc}
\toprule \hline
\textbf{Parameter} & \textbf{Prior Range} & \textbf{Units} \\
\hline
Right Ascension, RA              & [0, $2\pi$]            & rad \\
Declination, Dec                 & [$-\pi/2$, $\pi/2$]    & rad \\
Geocentric time, $t_c$           & [--0.5, 0.5]            & s \\
Luminosity distance, $d_L$       & [1, 10000]             & Mpc \\
Inclination angle, $\theta_{\rm JN}$ & [0, $\pi$]             & rad \\
Polarization angle, $\psi$       & [0, $\pi$]             & rad \\
Coalescence phase, $\phi_c$      & [0, $2\pi$]            & rad \\
\hline
\end{tabular}
\label{tab:prior_ranges}
\end{table}

Given the extremely high signal-to-noise ratios (Figure~\ref{fig:networksnrs}) and the long signal durations in some of our low-mass injections, nested samplers such as \texttt{dynesty} \citep{dynesty2020} or \texttt{nestle} face computational challenges in efficiently exploring the resulting sharply peaked posterior distributions. We therefore adopt the Fisher Information Matrix (FIM) approximation \citep{2008vallisneri}, which approximates the log-likelihood around the maximum-likelihood point to produce an analytic Gaussian approximation to the posterior. This method is well-justified in the high-SNR limit.
From this, we generate posterior samples that can be used to construct credible regions. 


\subsection{Localization Volumes} \label{localizationvolume}
For each simulated GW event, we compute the 3D localization volume by combining the sky localization area with luminosity distance constraints derived from the FIM-based posterior distributions. We begin by extracting the posterior samples for RA, Dec, and luminosity distance $d_L$. To estimate the sky localization area, we apply a 2D kernel density estimate (KDE) \citep{kde1, kde2} to the RA and Dec samples, evaluate the density on a HEALPix grid \citep{healpix} with resolution parameter $\texttt{nside} = 2048$, and identify the smallest set of pixels enclosing 50\% and 90\% of the posterior probability. Figure~\ref{fig:skymap} shows an example of the 2D localization areas of a 50~\Msun~+~50~\Msun~ injection at 1000 Mpc. 
We extract the central credible interval from the $d_L$ posterior distribution, convert the lower and upper bounds to redshift using the assumed cosmology, and compute the comoving shell volume between $z_\mathrm{min}$ and $z_\mathrm{max}$. The final 3D localization volume is obtained by multiplying this shell volume by the fractional solid angle subtended by the sky localization region. The 50\% ($V_{50}$) and 90\% ($V_{90}$) localization volumes are then used for further analyses.

\subsection{Galaxy Mass Weighting Across BBH Formation Channels} \label{gsmf_mod}
Several formation channels have been proposed for BBHs (see \citealt{Mapelli2020} for a detailed review): BBH mergers can be the result of isolated binary evolution via common envelope \citep{Bethe_1998, 1998zwart, 2002belczynski, 2008belczynski, Belczynski_2016, dvorkin2016, 2018dvorkin, eldridge_stanway_2016, mapelli2017, 2019mapelli, stevenson_2017, 2018kruckow, Spera_2019, bel2020, klencki2021, 2021olejak, tanikawa2021}, stable mass transfer \citep{giacobbo2018, 2019neijssel, 2021bavera, Gallegos-Garcia_2021, Shao_2021}, or chemically homogeneous evolution \citep{2016mink_mandel, 2016mandel_mink, 2016marchant, 2020bus, 2021riley}. Alternatively, BBHs can form dynamically in triples \citep[e.g.][]{2017antonini, Silsbee_2017, sedda2021}, multiples \citep[e.g.][]{fragione_2019, Hamers_2020}, young stellar clusters \citep[e.g.][]{2010banerjee, mapelli2016, 2017banerjee, 2020kuma}, globular clusters (GCs) \citep{2000port, 2013tanikawa, 2014samsing, 2016rodri, 2017askar, fragione2018, hong2018, 2019choksi}, and nuclear star clusters \citep[e.g.][]{2020arcasedda, Fragione_2020}. 

These disparate formation channels may result in markedly different formation environments; however, the delay time between the formation and merger also means that the formation environment and merger environment are not necessarily identical \citep[e.g.][]{Mapelli2020}. As a first-order approximation, we can assume that the likelihood of a galaxy hosting a BBH merger is proportional to its stellar mass. In the local Universe ($\lesssim$1000 Mpc), most of the stellar mass is concentrated around \Mstar, the characteristic break in the galaxy stellar mass function (see below). However, given the strong relationship between the galaxy stellar mass and metallicity \citep{2004tremonti}, systems formed preferentially at low metallicity should also form in low mass galaxies. Alternatively, since the bulge mass and the number of GCs per unit stellar mass increases for more massive galaxies, those formed via dynamical interactions may be formed preferentially in more massive galaxies. We do not attempt to model this fully here, but consider characteristic (if simplistic) models that may capture the range of plausible behaviour.


To implement this approach, we begin with the galaxy stellar mass function (GSMF) as described by the double Schechter function \citep{1976schechter}:
\begin{equation} \label{schechter}
    \displaystyle \Phi(M) \rm{d}M = \left[\phi_1\left(\frac{M}{M_*}\right)^{\alpha_1}+\phi_2\left(\frac{M}{M_*}\right)^{\alpha_2}\right] e^{-M / M_*} \frac{\rm{d}M}{M_*},
\end{equation}
where $M$ is the stellar mass of the galaxy, $\Phi(M)dM$ is the number density of galaxies per unit stellar mass in the range $(M, M+dM)$, \Mstar~is the characteristic stellar mass which represents the transition between the power-law and exponential regimes, $\phi_1$ and $\phi_2$ are the normalization constants for the high-mass and low-mass ends respectively, and $\alpha_1$ and $\alpha_2$ are the power law slopes of the high-mass and low-mass ends, respectively. To compute the number density of galaxies in a specific mass range $(M_1, M_2)$, we integrate Equation~\ref{schechter} over those limits. Observational values for the Schechter parameters ($M_*$, $\phi_1$, $\phi_2$, $\alpha_1$, $\alpha_2$) are drawn from various galaxy surveys, including GAMA \citep{2012baldry}, SDSS DR7 \citep{2016weigel}, and high-redshift samples from the Hubble Space Telescope (HST) and other ground-based observatories \citep{2021hst} (see Table~\ref{tab:galaxy_mass_function}).

\begin{table}
  \centering
  \small
  \caption{GSMF parameters for galaxies in various surveys over different redshifts.}
  \label{tab:galaxy_mass_function}
  \begin{tabular}{@{}lcccccc@{}}
    \toprule
    Ref. & $z$ & $\log(M_*/M_\odot)$ & $\phi_1$ & $\phi_2$ & $\alpha_1$ & $\alpha_2$ \\
    & & & \multicolumn{2}{c}{($10^{-3}$ Mpc$^{-3}$)} & & \\
    \midrule
    B12 & < 0.06 & 10.66 & 3.96 & 0.79 & -0.35 & -1.47 \\
    \addlinespace
    W16 & 0.02 - 0.06 & 10.79 & 0.49 & 0.98 & -1.69 & -0.79 \\
    \addlinespace
    M21 & 0.25 - 0.75 & 10.64 & 2.34 & 0.78 & 0.25 & -1.49 \\
    & 0.75 - 1.25 & 10.51 & 2.14 & 0.85 & 0.08 & -1.49 \\
    & 1.25 - 1.75 & 10.54 & 1.48 & 0.48 & -0.07 & -1.60 \\
    & 1.75 - 2.25 & 10.56 & 0.89 & 0.31 & -0.06 & -1.63 \\
    & 2.25 - 2.75 & 10.55 & 0.53 & 0.32 & 0.02 & -1.66 \\
    & 2.75 - 3.75 & 10.64 & 0.08 & 0.18 & 0.35 & -1.76 \\
    \bottomrule
  \end{tabular}
  \begin{tablenotes}
    \small
    \item Ref.: B12 - \cite{2012baldry}, W16 - \cite{2016weigel}, M21 - \cite{2021hst}
    \item Surveys: GAMA (B12), SDSS DR7 (W16), and HST combined with ground-based surveys \citep{2021hst}
  \end{tablenotes}
\end{table}

While the GSMF captures the overall stellar mass distribution of galaxies, different BBH formation channels weigh this distribution differently based on their host environment preferences. We first consider the isolated formation channel. \citet{2022_santoli} provide empirical fits for the BBH merger rate per galaxy per year, $n_{\rm iso}$\footnote{$\rm log(n_{\rm iso}/Myr^{-1}) = a + b~log(M/M_\odot)$}, across a range of stellar masses and redshifts\footnote{We assume the model corresponding to common-envelope parameter $\alpha$=5 (see also \citealt{2021zevin_e_bbhform} and \citealt{2025colloms}); $\alpha$ quantifies the efficiency with which orbital energy is used to expel the envelope of a star when the two stars in a binary interact closely.}. These fits, based on population synthesis models, incorporate either the galaxy mass-metallicity relation (MZR) or the fundamental metallicity relation (FMR). As both yield similar BBH merger rate densities for $z < 1$, we adopt the MZR-based values for our analysis (see their table 2 and figure 13).

We then normalize the BBH merger rate such that $n_{\rm iso} \in [0, 1]$ and define an efficiency-weighted GSMF:
\begin{equation} \label{phi_iso}
\displaystyle \Phi^{\rm iso}(M, z) = n_{\rm iso}(M)~ \Phi(M, z).
\end{equation}

\begin{figure}
    \centering
    \includegraphics[width=\linewidth]{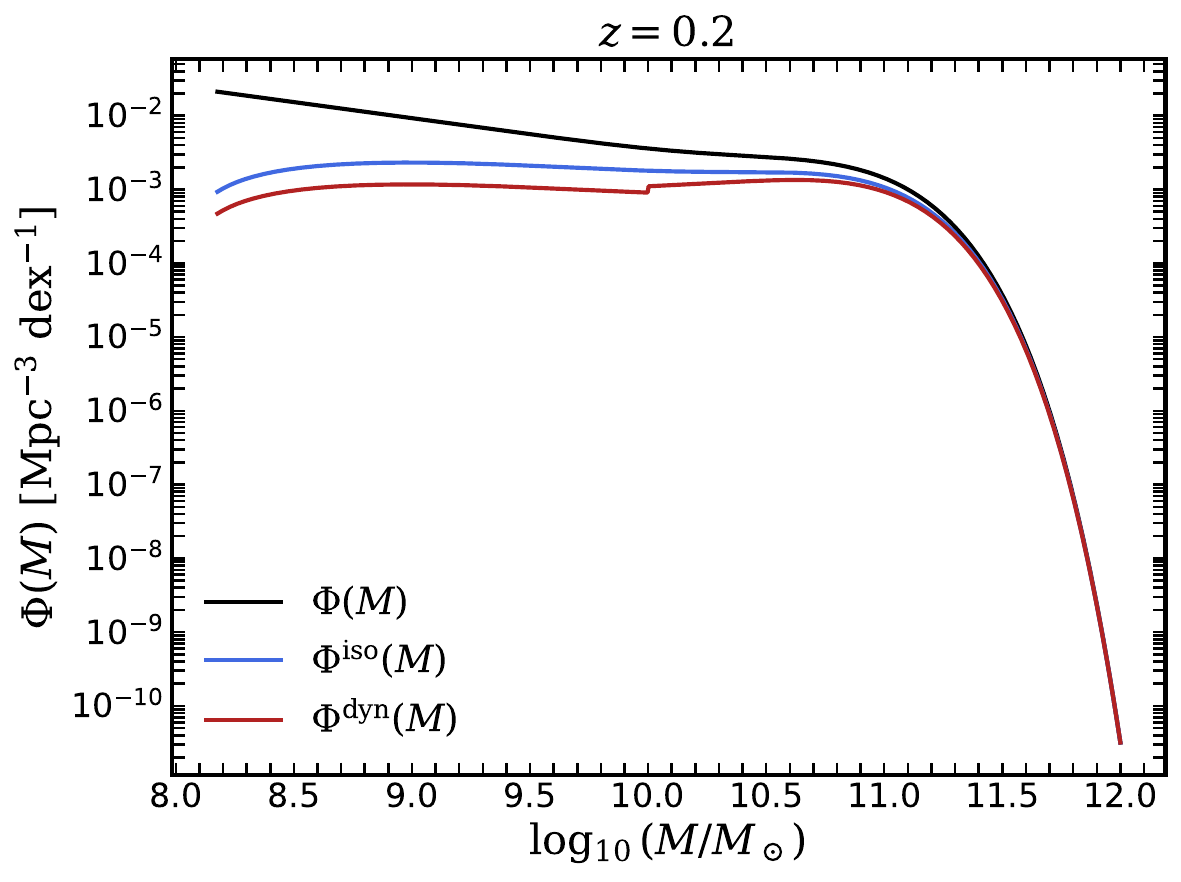}
    \caption{Weighted and unweighted GSMFs at $z = 0.2$; The \textit{black} curve shows the original GSMF $\Phi(M)$ (Equation~\ref{schechter}), modeled using a redshift-interpolated double Schechter function. The \textit{blue} curve, $\Phi^{\rm iso}(M)$ (Equation~\ref{phi_iso}), represents the mass function weighted by a BBH merger efficiency model for the isolated formation channel, normalized over stellar mass. The \textit{red} curve, $\Phi^{\rm dyn}(M)$ (Equation~\ref{phi_dyn}), is weighted by a globular cluster (GC) scaling relation to represent dynamical BBH formation.}
    \label{fig:phi_comparison}
\end{figure}

To model the dynamical channel, we assume that BBHs originate from interactions in GCs. In this case, the BBH formation efficiency would follow an empirical power-law relation between the galaxy stellar mass and the number of GCs. To account for this, we adopt a stellar mass–GC scaling relation to weigh the GSMF, thereby constructing a dynamical efficiency-weighted GSMF. The number of GCs per galaxy with $M > 10^{10}$ \Msun~is well described by a power-law scaling relation \citep{2013harris, 2014harris, 2015zaritsky}:
\begin{equation} \label{GC}
\displaystyle n_{\rm GC}(M) = N_0 \left( \frac{M}{M_0} \right)^\delta,
\end{equation}
where $M$ is the stellar mass of a galaxy, $M_0$ is a reference mass, $N_0$ is a normalization constant, and $\delta$ is the power-law slope. For $M > 10^{10}~\rm{M_\odot}$, we adopt $\log N_0 = 2.924$, $\log (M_0/\rm{M_\odot}) = 11.2$, and $\delta \approx 1$ \citep{2013harris}. For lower-mass galaxies ($10^8 \leq M / \rm{M_\odot} \leq 10^{10}$), the relation flattens, with $\delta = 0.365$, $\log N_0 = 1.274$, and $\log (M_0/\rm{M_\odot}) = 9.2$. As before, we normalize $n_{\rm GC}(M)$ such that $n_{\rm GC}(M) \in [0, 1]$ and define the dynamical efficiency-weighted GSMF:
\begin{equation} \label{phi_dyn}
\displaystyle \Phi^{\rm dyn}(M, z) = n_{\rm GC}(M) ~ \Phi(M, z).
\end{equation}
Figure~\ref{fig:phi_comparison} illustrates the original GSMF $\rm \Phi(M)$, and the weighted GSMFs: $\rm \Phi^{iso}(M)$ and $\rm \Phi^{dyn}(M)$ at $z = 0.2$. The isolated BBH formation channel yields a broad and uniform distribution across galaxy stellar masses, whereas the dynamical channel shows a relative preference for higher-mass galaxies, reflecting the scaling of the number of GCs with galaxy mass.

\subsection{Theoretical Localization Volumes} \label{theoreticalvol}
To evaluate the prospects for uniquely identifying the host of a BBH merger, we compare the simulated GW localization volume to the expected number density of galaxies within that same volume. In particular, we evaluate the minimum comoving volume that must be achieved in order to uniquely associate a galaxy of stellar mass \Mstar~with the BBH merger. This provides a theoretical upper bound on the localization volume necessary for host identification, under the assumption of a specific BBH formation channel.

We assume that galaxies are distributed homogeneously and isotropically in space, following the cosmological principle. While true on large scales and a standard approximation  \citep[e.g.][]{dodelson1997}, we note that on small scales, this approximation necessarily breaks down. True localization volumes are likely not within this limit (at least not when they are sufficiently small to yield host identifications); hence, this approach is necessarily an approximation. 
Given a GSMF, $\Phi (M, z)$ (Equation~\ref{schechter}), we compute the expected number of galaxies with stellar masses $\geq M$ contained within a comoving volume $V$ by assuming Poisson statistics. The minimum comoving volume required to contain, on average, $\lambda$ such galaxies is then
\begin{equation}
V_{\min}(\geq M, z)
=
\frac{\lambda}{\displaystyle\int_{\log_{10} M}^{\infty} \Phi(M', z)~d\log_{10} M'} ,
\label{vmin_eqn}
\end{equation}
where the denominator represents the cumulative comoving number density of all galaxies with stellar masses $\geq M$. This cumulative form reflects the fact that we do not know \emph{a priori} which galaxy within a localization volume is the true host. 

\begin{figure*}
    \centering
    \includegraphics[width=\linewidth]{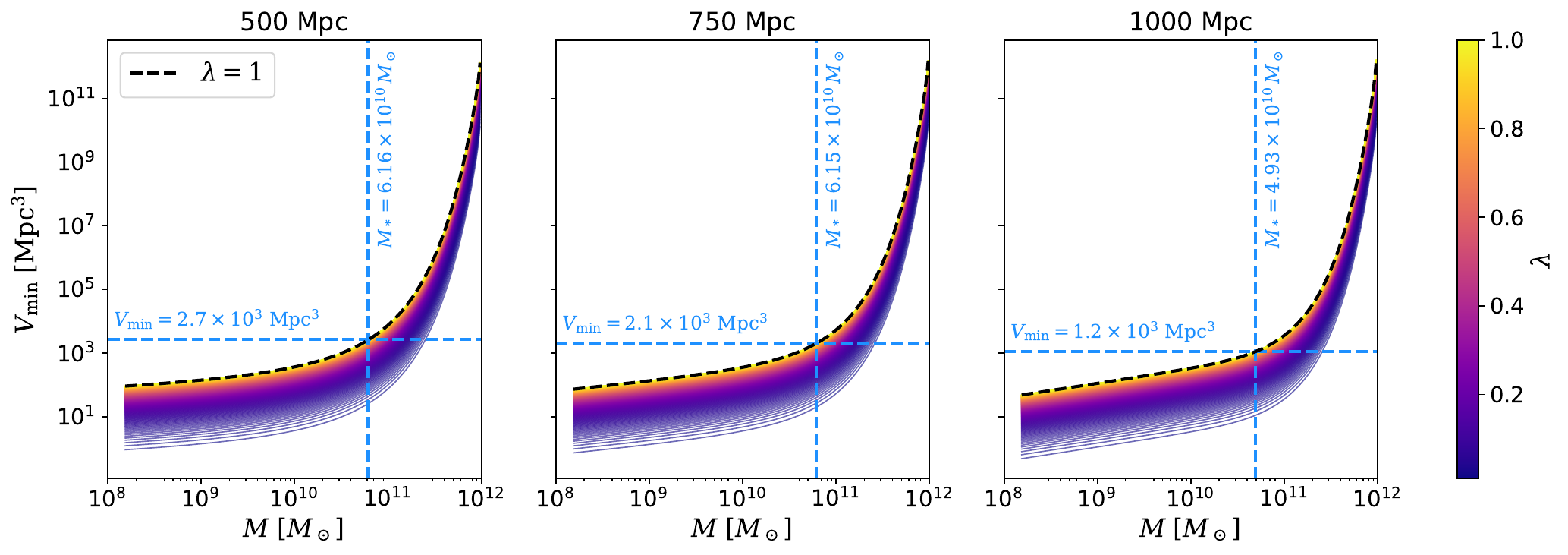}
    \caption{Minimum comoving volumes $V_{\min}$ required to contain, on average, one galaxy of mass $M$ as a function of galaxy stellar mass and redshift. Each panel corresponds to a different injected host at distances of 500 Mpc, 750 Mpc, and 1000 Mpc. Coloured curves show $V_{\min}(M)$ (Equation~\ref{vmin_eqn}) scaled by different values of $\lambda$. The black dashed line denotes the fiducial case of $\lambda = 1$. The vertical blue dashed line marks the mass of the injected \Mstar~galaxy (Grid I, Table~\ref{tab:injected_hosts}), while the horizontal blue dashed line indicates the corresponding $V_{\min}$ value at that mass. All three curves assume the redshift-dependent GSMF $\Phi (M, z)$ (Equation~\ref{schechter}).}
    \label{fig:vmin_comparison}
\end{figure*}

A localization volume equal to $V_{\min} (\lambda = 1)$ corresponds to an expectation value of, on average, one galaxy of mass $M_*$; volumes smaller than this may still contain such a galaxy, but the average number falls below unity (Figure~\ref{fig:vmin_comparison}).



To account for different astrophysical channels of BBH formation, we compute analogous minimum comoving volumes using the modified GSMFs introduced earlier in Section~\ref{gsmf_mod}. For the isolated and dynamical channels, Equation~\ref{vmin_eqn} is evaluated using $\Phi^{\rm iso} (M, z)$ (Equation~\ref{phi_iso}) and $\Phi^{\rm dyn} (M, z)$ (Equation~\ref{phi_dyn}), resulting in corresponding minimum localization volumes, $V^{\rm iso}_{\min} (M, z)$ and $V^{\rm dyn}_{\min} (M, z)$, respectively. Each of these volumes - $V_{\min}$, $V^{\rm iso}_{\min}$, and $V^{\rm dyn}_{\min}$ - can be interpreted as the minimum localization volume required to statistically isolate, on average, a single galaxy of stellar mass \Mstar~($\lambda = 1$) under the assumption of the corresponding BBH formation channel. We note, importantly, that $\lambda=1$ is the point at which we expect one galaxy within the localization volume at random. Hence, it is not a volume at which we can realistically claim (at least not with high confidence) to be able to identify the host. We consider this further in Section~\ref{hg_metrics}.  

\begin{table}
\centering
\caption{Minimum volume required to observe, on average, at least 1 galaxy of mass M (equal to the injected host masses at each corresponding redshift, see Table~\ref{tab:injected_hosts}) ($V_{\min}$) for each BBH formation channel. Units are in Mpc$^3$.}
\begin{tabular}{lccc}
\toprule
\textbf{Channel} & \textbf{500~Mpc} & \textbf{750~Mpc} & \textbf{1000~Mpc} \\
\midrule
$V_{\min}$ & $2.7 \times 10^3$  & $2.1 \times 10^3$  & $1.2 \times 10^3$ \\
$V_{\min}^{\mathrm{iso}}$  & $3.6 \times 10^3$  & $2.7 \times 10^3$  & $1.5 \times 10^3$ \\
$V_{\min}^{\mathrm{dyn}}$ & $4.1 \times 10^3$  & $3.2 \times 10^3$  & $1.8 \times 10^3$ \\
\bottomrule
\end{tabular}
\label{tab:vmin_channels}
\end{table}

\subsubsection{Metallicity Dependence} \label{metallicity}
Several binary population synthesis models predict that BBH formation via isolated binary evolution is strongly dependent on progenitor metallicity \citep[e.g.][]{2010bel, eldridge_stanway_2016}, with lower-metallicity environments favouring the formation of more massive BHs and higher merger rates \citep[e.g.][]{giacobbo2018, Spera_2019}. In contrast, dynamical formation channels are largely insensitive to metallicity \citep[e.g.][]{2022mapelli}. This distinction presents a potential way to distinguish BBH formation channels. 
Analysis by \citet{2021zevin_e_bbhform} (see also \citealt{2021bouff}) indicates that no single formation pathway accounts for more than $\sim$70\% of the BBHs detected in GWTC-2.0 and GWTC-2.1 \citep{gwtc_2a, gwtc_2b}. Recent population synthesis studies suggest that the isolated channel may contribute a larger fraction than the dynamical channel \citep{barbar2025}. 

To explore how metallicity may influence our analysis, we adopt a representative metallicity threshold of $12 + \rm{log (O/H)} = 8.3$, commonly found as a threshold for the occurrence of long gamma-ray bursts (LGRBs) \citep[e.g.][]{2017graham}. The relation between galaxy stellar mass and metallicity is captured by the MZR relation of \citet{2014zahid}, valid for $z \lesssim 1.6$ and parameterized as:
\begin{equation} \label{zahid}
    \displaystyle \rm 12 + log (O/H) = Z_0 + log \left[1 - exp \left(- \left[ \frac{M_*}{M_0} \right]^\gamma   \right) \right],
\end{equation}
where $Z_0$ is the saturation metallicity \citep[e.g.][]{2011moustakas}; $M_0$ is the characteristic turnover galaxy mass above which the metallicity asymptotically approaches the upper metallicity limit $Z_0 = 9.102$ and $\gamma = 0.513$ is the power-law index (values taken from table~2 in \citealt{2014zahid}). The MZR flattens at high galaxy stellar masses at $z=0.2$. Using this relation, we find that this threshold corresponds to a galaxy stellar mass of log(M/\Msun) $ = \rm M_Z$ = 7.97. We then treat this mass as a lower threshold and compute the number density of galaxies with stellar mass $\geq \rm M_Z$, using a method analogous to that described in Section~\ref{theoreticalvol}. This yields $\Phi^Z = 1.84 \times 10^{-2}~\rm{Mpc^{-3}}$, and an associated minimum localization volume of $V^Z_{\min} = 54.42~\rm{Mpc^3}$.

\section{Prescriptions for Host Galaxy identification for BBH mergers} \label{hg_metrics}

\subsection{Masses within the GW Localization Volumes} \label{theoretical_mass}

We compare the injected host to the total stellar mass in galaxies within the GW localization volume. We compute the total stellar mass enclosed in two ways: a theoretical estimate based on the GSMF, and an observational estimate obtained by summing the stellar masses of galaxies in galaxy catalogues within the localization region, although we note that, at least at the current time, the latter approach is of very limited value because of galaxy catalogue incompleteness. 

\subsubsection{Theoretical Galaxy Stellar Mass}
To estimate the total stellar mass theoretically enclosed within a GW localization region, we proceed as follows. We compute the stellar mass density, $\rm \rho_\star(z)$ [$\rm {M_\odot~Mpc^{-3} }$] by computing the mass-weighted integral of the GSMF, $\Phi(M)$, over a stellar mass range of ($M_{\min}, M_{\max} = 10^8$~\Msun, $10^{12}$~\Msun):
\begin{equation} \label{rho_star}
    \displaystyle \rho_\star(z) = \int_{M_{\text{min}}}^{M_{\text{max}}} M~\Phi(M)~\mathrm{d}M.
\end{equation}
The total expected stellar mass enclosed within the GW localization region is $\rm M_{\Phi}(z) = \rho_\star~V_{\text{comoving}}$. This formalism enables the computation of the theoretical stellar mass contained within the $V_{50}$ and $V_{90}$ credible regions of each injection. 

\subsubsection{Observed Galaxy Stellar Mass}
We determine the total observed mass enclosed within the GW localization volume by first performing a 3D crossmatch of the GW localization volumes with the NED-LVS galaxy catalogue, and then defining:
\begin{equation} \label{obs_mass_eqn}
    \rm  M_{\text{obs}} =\sum_{i=1}^{N} M_{\mathrm{candidate},i},
\end{equation}
where $M_{\mathrm{candidate}, i}$ denotes the stellar mass of the $\rm i^{th}$ galaxy identified within the localization volume, and N is the total number of crossmatched candidates from the NED-LVS galaxy catalogue.

\subsubsection{Mass Fractions} \label{massratios}
To quantify the significance of the injected \Mstar~host relative to the total galaxy stellar mass within the GW localization volume, we define two complementary dimensionless mass fractions:

\begin{enumerate}
    \item \textbf{$\rm \mathbb{M}_{*, \Phi}$:} We compare the stellar mass of the injected \Mstar~host to the total theoretical mass enclosed within the GW localization volume by computing the theoretical mass fraction, $\rm \mathbb{M}_{*, \Phi} = M_{*}/{M_{\Phi}}$.

    \item \textbf{$\rm \mathbb{M}_{*, obs}$:} Similarly, we compute an observed mass fraction, $\rm \mathbb{M}_{*, obs} = M_{*}/{M_{obs}}$, where $M_{\rm obs}$ is the total stellar mass of all candidate hosts identified within the GW localization volume. 
\end{enumerate}

\subsection{Probability of Chance Alignment} \label{pchance}
A common way to assess the significance of a candidate host association is the probability of chance alignment, $p_c$. This quantifies the likelihood that a galaxy of a certain brightness falls at a certain proximity to the source location at random. It is calculated through a combination of the offset and magnitude of the putative host based on the 2D distribution of galaxies on the sky. In particular, \citet{2002bloom} define $P_{c} = 1 - \exp{(-\eta_i)}$. Here $\eta_i$ is the expected number of galaxies as bright as or brighter than the candidate host within a certain search region. In the traditional 2D case, this is given by $\eta_i = \pi r^2 \sigma_i$, where $r$ is the angular separation from the transient and $\sigma_i$ is the surface density of galaxies as bright as or brighter than the host. 

In the context of GW follow-up, the projected 2D density of galaxies across the sky becomes extremely high due to the large localization areas. However, as the measurement of the luminosity distance is also available, this information should be used as well. The total number of galaxies within the 3D localization volume (i.e., within the luminosity distance and sky position uncertainty) may be tractable.

We therefore define $\eta_i$ to account for the number of galaxies with stellar mass equal to or greater than that of the candidate host, $M_{\rm host}$, within a comoving volume $V$:
\begin{equation} \label{eta_eqn}
    \displaystyle \eta_i (\geq M_{\rm host}) = V \int_{M_{\rm host}}^{\infty} \Phi(M) dM
\end{equation}
where $\Phi(M)$ is the GSMF (Equation~\ref{schechter}). Here, we consider both the $V_{50}$ and $V_{90}$ credible volumes for $V$, and also the weighted GSMFs corresponding to the different BBH formation channels, $\Phi^{\rm iso}_{\min}$ (Equation~\ref{phi_iso}) and $\Phi^{\rm dyn}_{\min}$ (Equation~\ref{phi_dyn}). 

Alternatively, a luminosity-based formulation can be used:
\begin{equation} \label{eta_lum_eqn}
\displaystyle \eta_i(\geq L_{\rm host}) = V \int_{L_{\rm host}}^{\infty} \Phi(L) \, dL,
\end{equation}
where $\Phi(L)$ is the galaxy luminosity function and $L_{\rm host}$ is the luminosity of the candidate host.

\section{Results} \label{results}


\subsection{Grid I: Galaxy Catalogue Injections}
\subsubsection{Localization Volumes}
\begin{figure*}
    \centering
    \includegraphics[width=\linewidth]{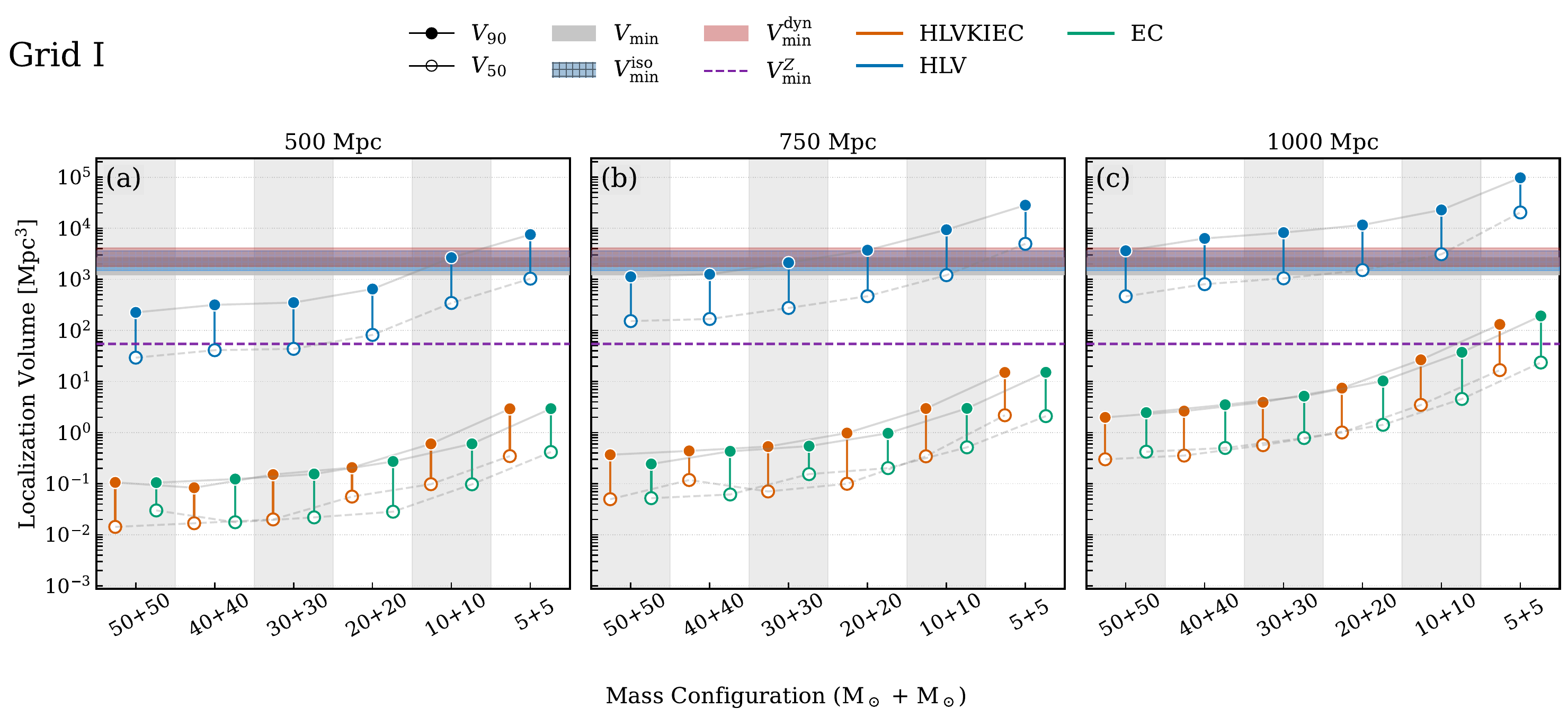}
    \caption{
    Localization volumes for simulated BBH mergers in Grid I, at luminosity distances of 500, 750, and 1000 Mpc \textit{(Panels a–c)}. For each mass configuration and network: HLVKIEC \textit{(red)}, HLV \textit{(blue)}, EC \textit{(green)}, we plot $V_{50}$ \textit{(hollow marker)} and $V_{90}$ \textit{(filled marker)}, with a vertical line connecting the two. Horizontal shaded bands indicate the minimum comoving volume required to contain, on average, one galaxy of mass \Mstar~or higher under three different BBH formation channel assumptions: $V_{\min}$ (\textit{gray}, Equation~\ref{vmin_eqn}), $V_{\min}^{\mathrm{iso}}$ (\textit{blue}), and $V_{\min}^{\mathrm{dyn}}$ (\textit{red}). These threshold comoving volumes were computed within a redshift range of $z = 0.12$ and $z = 0.23$, corresponding to the injected~\Mstar~hosts (Table~\ref{tab:vmin_channels}). The dashed purple horizontal line corresponds to the metallicity-dependent minimum comoving volume, $V^Z_{\min}$ = 370.70 $\rm Mpc^3$, as calculated in Section~\ref{metallicity}.
    }
    \label{fig:vol_comp_I}
\end{figure*}

Figure~\ref{fig:vol_comp_I} presents the 3D localization volumes, quantified by the $V_{50}$ and $V_{90}$ credible regions, for each BBH injection in Grid I (Table~\ref{tab:injected_hosts}), as inferred for the three GW detector networks we consider: HLVKIEC, HLV, and EC. Consistent with expectations, the HLVKIEC and EC networks yield significantly smaller localization volumes than HLV, due to their enhanced detector sensitivity and correspondingly higher SNR values for the CBCs. At fixed network and distance, localization volumes increase with decreasing total mass of the binary due to the correlation between SNR and CBC masses (Figure~\ref{fig:networksnrs}). For high-mass binaries, such as 50+50~\Msun, $V_{90}^{\rm HLVKIEC}\sim V_{90}^{\rm EC} \sim 10^{-1}~\rm{Mpc}^3$ at 500~Mpc, and $V_{90}^{\rm HLVKIEC}\sim V_{90}^{\rm EC}\sim 10~\rm{Mpc}^3$ at 1000~Mpc. In contrast, the HLV network yields much poorer localization, with $V_{90}^{\rm HLV} > 10^2~\rm{Mpc}^3$ across all the distances and mass configurations. 

The horizontal shaded bands in Figure~\ref{fig:vol_comp_I} indicate the theoretical minimum comoving volumes required to contain, on average, one galaxy of mass $M_*$ or higher, under the assumption of different BBH formation channels as defined in Section~\ref{theoreticalvol}, along with the metallicity-based threshold $V^Z_{\min}$ that we defined in Section~\ref{metallicity}. Across both 50\% and 90\% credible volumes, the EC and HLVKIEC networks demonstrate strong localization performance relative to the threshold volumes, with localization volumes consistently lower than the thresholds across all mass configurations and distances; only at the lowest mass configuration at 1000 Mpc, $\rm V^{HLVKIEC}_{90}$ and $\rm V^{EC}_{90}$ are above $V^Z_{\min}$. 




\subsubsection{Mass Fractions}
\begin{figure*}
    \centering
    \includegraphics[width=\linewidth]{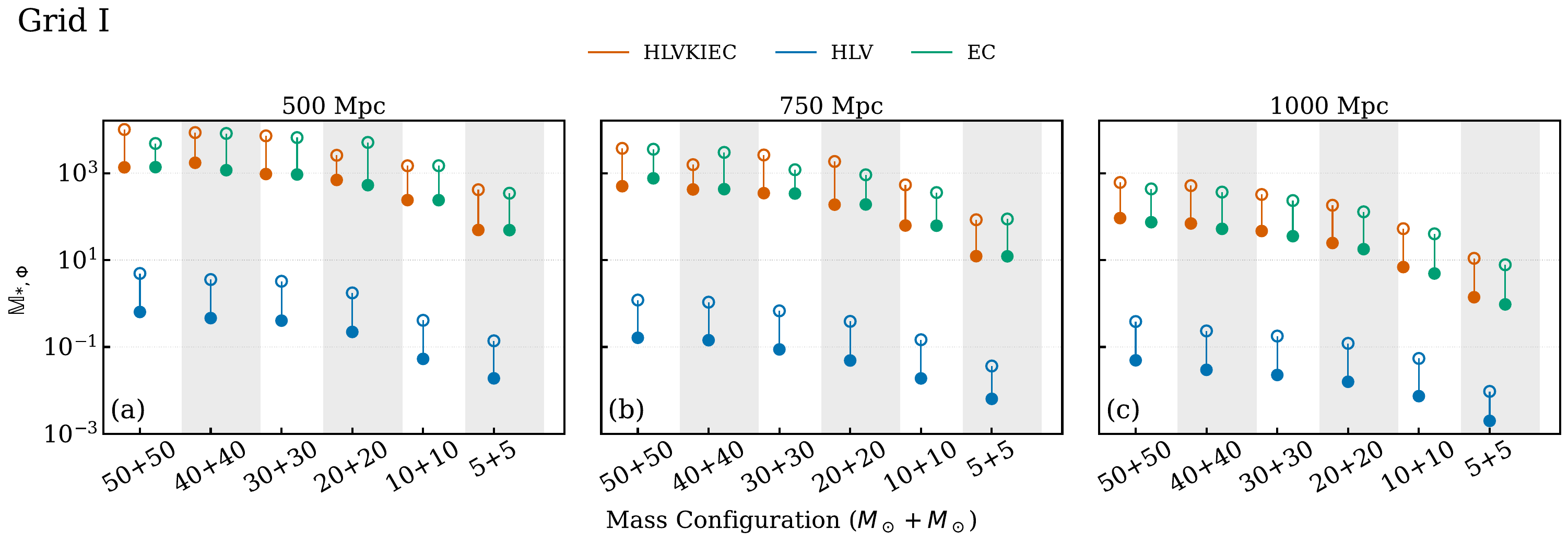}
    \caption{Theoretical ($\rm \mathbb{M}_{host, \Phi}$) mass fractions (Section~\ref{theoretical_mass}) for simulated BBH mergers in Grid I at 500 Mpc, 750 Mpc and 1000 Mpc \textit{(Panels a - c)}. $\rm \mathbb{M}_{host, \Phi}$ is the ratio between the mass of the injected \Mstar~galaxy and the total theoretical mass enclosed within the GW localization volumes $V_{50}$ and $V_{90}$.Colours denote the GW detector networks: HLVKIEC \textit{(red)}, HLV \textit{(blue)}, EC \textit{(green)}. The \textit{hollow} markers correspond to values evaluated within the $V_{50}$ region and the \textit{filled} markers correspond to values evaluated within the $V_{90}$ region.}
    \label{fig:gridI_massratio}
\end{figure*}

Figure~\ref{fig:gridI_massratio} shows the theoretical mass fractions, $\rm \mathbb{M}_{*, \Phi}$, for each BBH injection in Grid I. Both HLVKIEC and EC networks yield values of $\sim 1$ - $10^{4}$ at all three distances, for values evaluated with both $V_{90}$ and $V_{50}$, with a systematic decline towards 1000 Mpc and towards low-mass BBHs; for HLV, $\rm \mathbb{M}_{*, \Phi}$ is a few orders of magnitudes smaller ($\sim 10^{-3} - 10$), reflecting the larger localization volumes in those cases. The corresponding observed mass fractions, $\rm \mathbb{M}_{*, obs}$, are discussed in Section~\ref{massfraction_galcat_incomp}.



\subsubsection{Chance Alignment}
\begin{figure*}
    \centering
    \includegraphics[width=\linewidth]{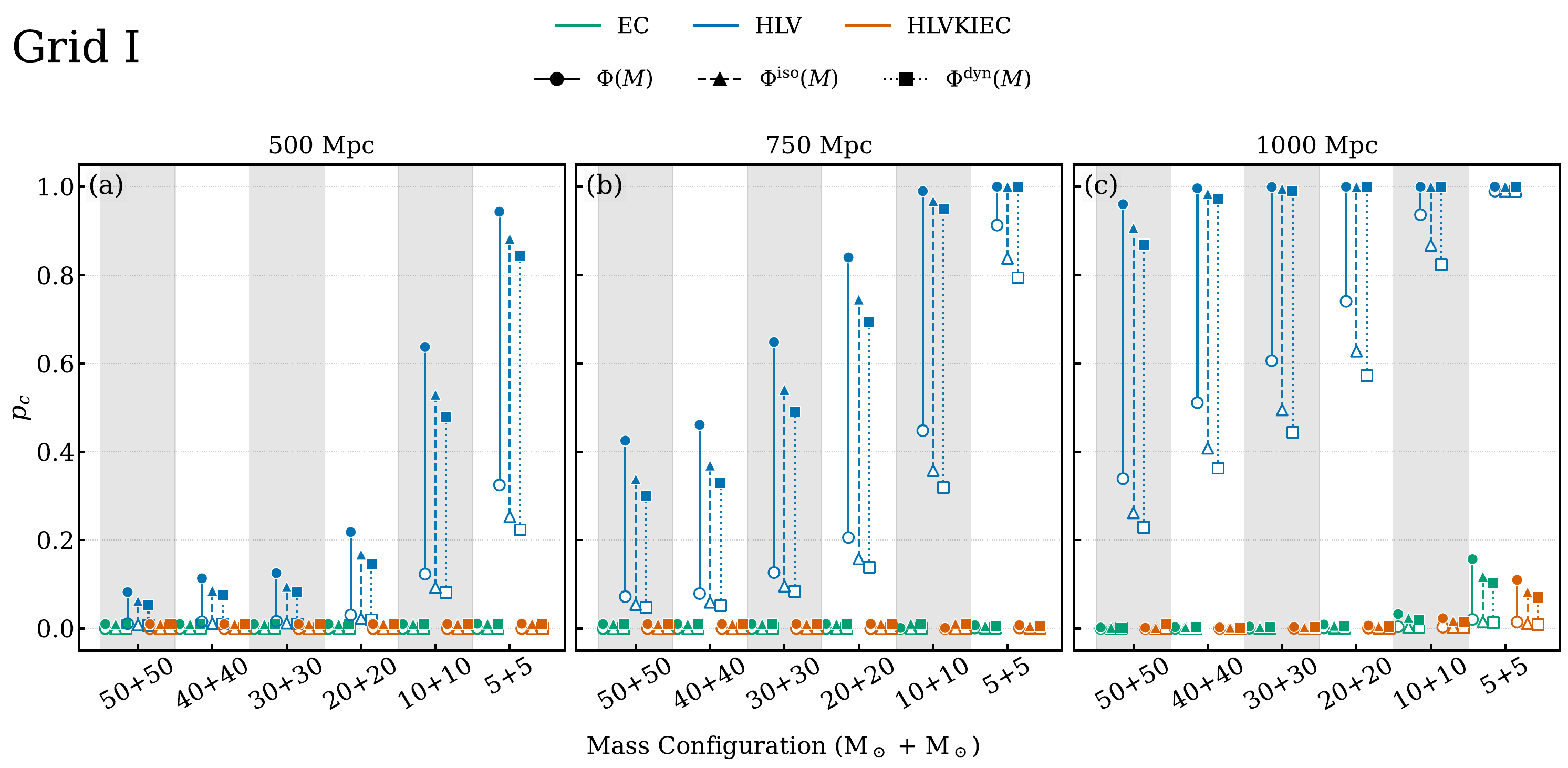}
    \caption{Probability of chance alignment, $p_c$, for Grid I BBH injections at 500 Mpc, 750 Mpc, and 1000 Mpc \textit{(Panels a-c)}. Colours denote the GW detector networks: HLVKIEC \textit{(red)}, HLV \textit{(blue)}, EC \textit{(green)}. The line style/marker encodes the different GSMFs across BBH formation channels used in Equation~\ref{eta_eqn}: $\Phi (M)$ (\textit{solid/circle}, Equation~\ref{schechter}), $\Phi^{\rm iso}(M)$ (\textit{dashed/triangle}, Equation~\ref{phi_iso}), $\Phi^{\rm dyn}(M)$ (\textit{dotted/square}, Equation~\ref{phi_dyn}). For each injection, the vertical segment spans the value at $V_{50}$ \textit{(hollow marker)} to the value at $V_{90}$ \textit{(filled marker)}. 
}
    \label{fig:pchance_I}
\end{figure*}

We evaluate the probability of change alignment, $p_c$ (Section~\ref{pchance}), as a diagnostic of host identification for the BBH injections in Grid I in Figure~\ref{fig:pchance_I}. At 500 Mpc, both EC and HLVKIEC consistently yield low chance alignment values, with $p_c < 0.01$ for all mass configurations and GSMFs. In contrast, HLV shows higher values, with $p_c(V_{90}) > 0.8$ for the 5+5~\Msun~injection. At 750 Mpc, the difference between the detector networks becomes more distinct. Both EC and HLVKIEC maintain $p_c < 0.01$ for all mass configurations. The HLV-only configuration performs comparatively poorly, with $p_c(V_{90}) > 0.2$ for the 50+50~\Msun~injection and $p_c(V_{90}) \sim 1$ at 5+5~\Msun. At 1000 Mpc, $p_c(V_{90}) \lesssim 0.01$ for all the HLVKIEC and EC injections up to 10+10~\Msun, and increases to $\sim 0.1$ at 5+5~\Msun. HLV returns $p_c \approx 1$ across all injections at 1000 Mpc.

\subsection{Grid II: Maximum \& Minimum Sky Sensitivity Injections}
\subsubsection{Localization Volumes}
\begin{figure*}
    \centering
    \includegraphics[width=\linewidth]{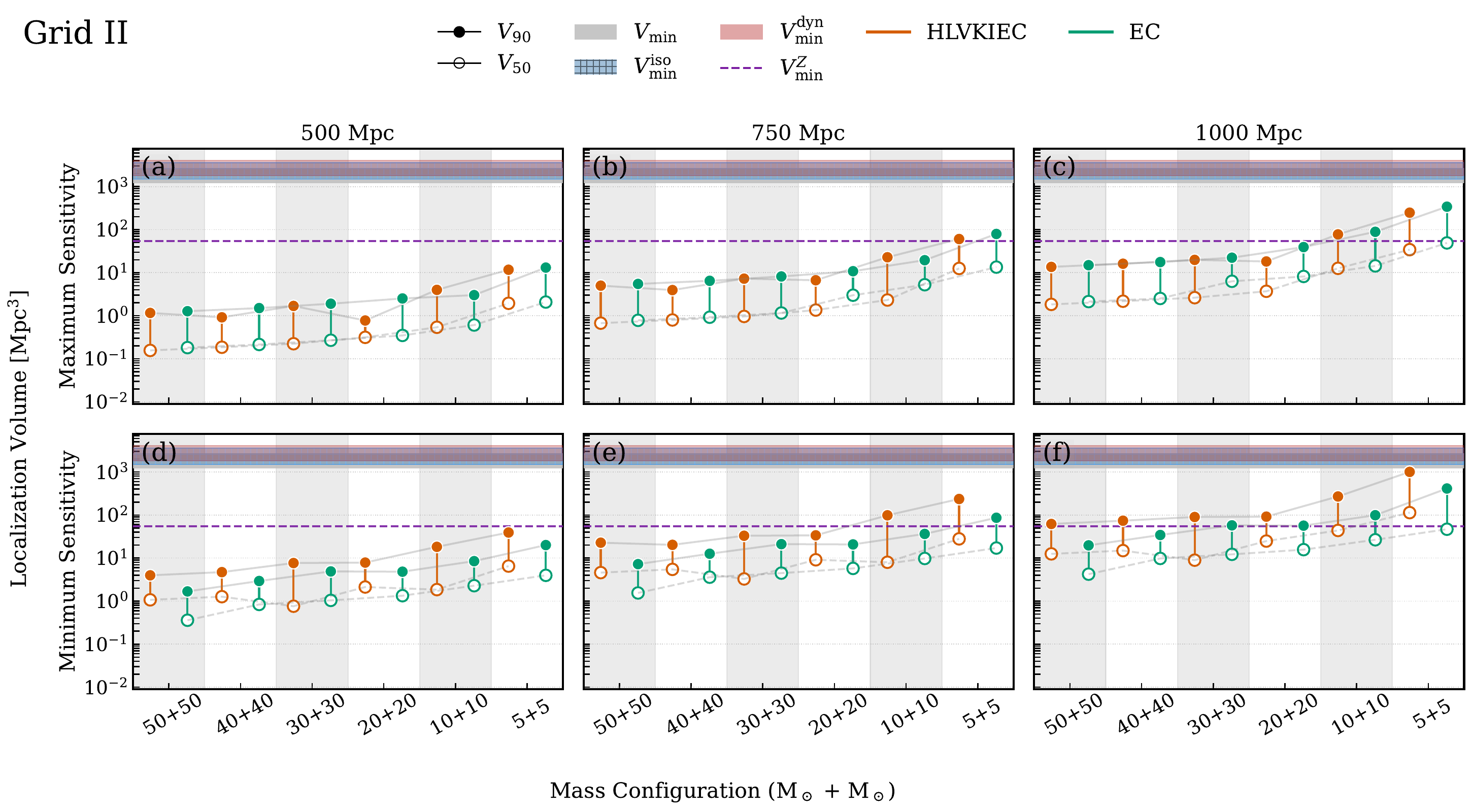}
    \caption{Localization volumes for simulated BBH mergers in Grid II, Top row \textit{(Panels a–c)} corresponds to the maximum sensitivity injections, and the bottom row \textit{(Panels d–f)} corresponds to the minimum sensitivity injections (Table~\ref{tab:bright_dark_points}). Columns correspond to luminosity distances of 500, 750, and 1000 Mpc. For each mass configuration and network: HLVKIEC \textit{(red)} and EC \textit{(green)}, we plot $V_{50}$ \textit{(hollow marker)} and $V_{90}$ \textit{(filled marker)}, with a vertical line connecting the two. Horizontal shaded bands indicate the minimum comoving volume required to contain, on average, one galaxy of stellar mass \Mstar~or of a higher mass, under three different BBH formation channel assumptions: $V_{\min}$ (\textit{gray}, Equation~\ref{vmin_eqn}), $V_{\min}^{\mathrm{iso}}$ (\textit{blue}), and $V_{\min}^{\mathrm{dyn}}$ (\textit{red}). These reference volumes were computed within a redshift range of $z = 0.12$ and $z = 0.23$, corresponding to the injected~\Mstar~hosts (Table~\ref{tab:vmin_channels}). The dashed purple horizontal line corresponds to the metallicity-dependent minimum comoving volume, $V^Z_{\min}$ = 370.73 $\rm Mpc^3$, as calculated in Section~\ref{metallicity}.
    }
    \label{fig:vol_comp_II}
\end{figure*}

Figure~\ref{fig:vol_comp_II} illustrates the impact of network directional sensitivity on localization performance by comparing the credible volumes, $V_{50}$ and $V_{90}$, for HLVKIEC and EC, at sky locations corresponding to maximum and minimum antenna responses (Table~\ref{tab:bright_dark_points}). As discussed in Section~\ref{inj_grids}, these sky positions were determined from the "bright" and "dark" regions of the antenna patterns of each detector network, and they approximately represent the most and least favourable configurations for network sensitivity, enabling us to probe the full range of localization outcomes for a fixed set of intrinsic source parameters. 

\textit{Panels (a-c)} in Figure~\ref{fig:vol_comp_II} show the localization volumes for the injected sources at the \emph{maximum sensitivity} points of each network (lime green circles in Figure~\ref{fig:antennapattern}, also see Table~\ref{tab:bright_dark_points}). In this best-case scenario, both networks localize all the injections across all the distances to volumes similar to or less than the threshold comoving volumes ($V_{\min}$, $V_{\min}^{\mathrm{iso}}$, $V_{\min}^{\mathrm{dyn}}$). For instance, a 50+50~\Msun~BBH merger at 500~Mpc has $V_{50}^{\rm HLVKIEC} \sim V_{50}^{\rm EC}$ $\rm \sim 10^{-1}~Mpc^3$. Across the distances, HLVKIEC shows slightly better localization performance than EC. At 500 Mpc, both networks have localization volumes $< V^Z_{\min}$; at higher distances, the lowest mass configurations are localized to volumes $\gtrsim V^Z_{\min}$. 

\textit{Panels (d-f)} present results from the \emph{minimum sensitivity} sky locations (white circles in Figure~\ref{fig:antennapattern}; also see Table~\ref{tab:bright_dark_points}). Localization performance slightly deteriorates in these regions, particularly for the low-mass systems and at 1000 Mpc. All the injections, except 5+5~\Msun~at 1000 Mpc, have localization volumes less than the  threshold comoving volumes. Notably, EC performs uniformly better than HLVKIEC in this case, we discuss this effect in detail in Section~\ref{networkcomparison}. At 500 Mpc, the localization volumes are $< V^Z_{\min}$ but at higher distances, heavier BBH systems have the localization volumes $\gtrsim V^Z_{\min}$, especially for $V^{\rm HLVKIEC}_{90}$. 


\subsubsection{Mass Fractions}
\begin{figure*}
    \centering
    \includegraphics[width=\linewidth]{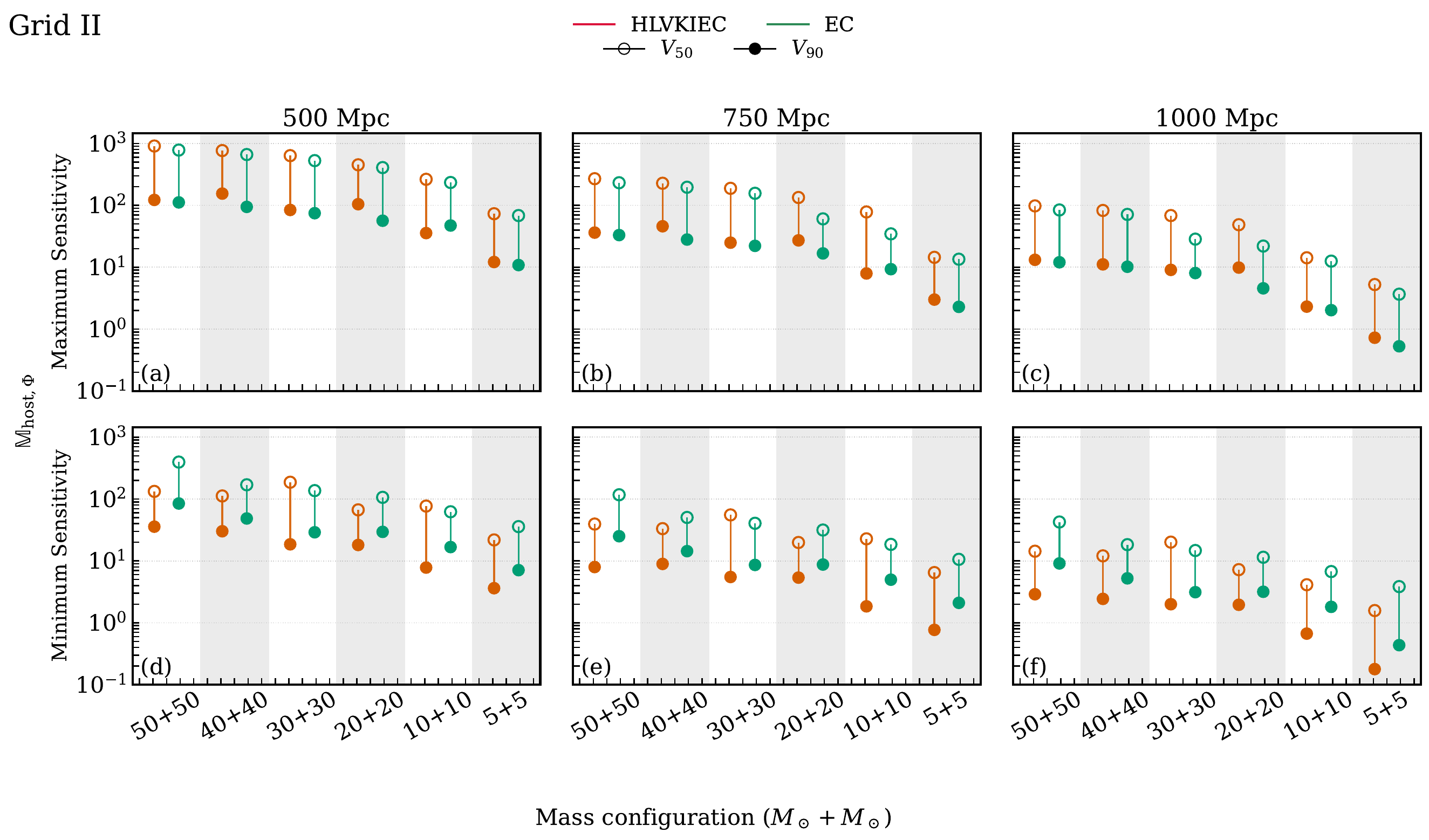}
    \caption{Theoretical mass fractions $\rm \mathbb{M}_{host, \Phi}$ for Grid II injections. Top row \textit{(Panels a - c)} corresponds to the maximum sensitivity injections, and the bottom row \textit{(Panels d - f)} corresponds to the minimum sensitivity injections (Table~\ref{tab:bright_dark_points}). Columns correspond to luminosity distances of 500, 750, and 1000 Mpc. For each mass configuration and network: HLVKIEC \textit{(red)} and EC \textit{(green)}, we plot the theoretical mass fraction calculated with $V_{90}$ \textit{(filled marker)} and $V_{50}$ \textit{(hollow marker)}. Fixed host masses of $M_{\rm host} = (6.16, 6.15, 4.93) \times 10^{10} M_\odot$ are adopted for 500, 750, and 1000 Mpc respectively (as in Grid I). 
    }
    \label{fig:gridII_massratio}
\end{figure*}

Figure~\ref{fig:gridII_massratio} shows the theoretical mass fractions, $\rm \mathbb{M}_{host, \Phi}$, calculated for the Grid II injections. For the \emph{maximum sensitivity} injections \textit{(Panels a-c)}, $\rm \mathbb{M}_{host, \Phi}$ is $\sim 10^2-10^3$ for most injections at 500 Mpc, falling to $\sim 10 - 10^2$ at 1000 Mpc; lighter BBH mergers have values $\sim 1$. For the \emph{minimum sensitivity} injections \textit{(Panels d-f)}, the fractions are lower by an order of magnitude across the grid, with most injections at 1000 Mpc, evaluated with $V_{90}$, having values $\approx 1$. Large values such as $\rm M_{host, \Phi} \sim 100$ imply that the volume contains less than one \Mstar~galaxy, i.e., $\rm M_\Phi \approx 0.01~M_{host}$. Observed mass fractions, $\rm \mathbb{M}_{host, obs}$, are not calculated for Grid II because the sky positions of the injected galaxy do not correspond to the NED-LVS galaxy catalogue.

\subsubsection{Chance Alignment}
\begin{figure*}
    \centering
    \includegraphics[width=\linewidth]{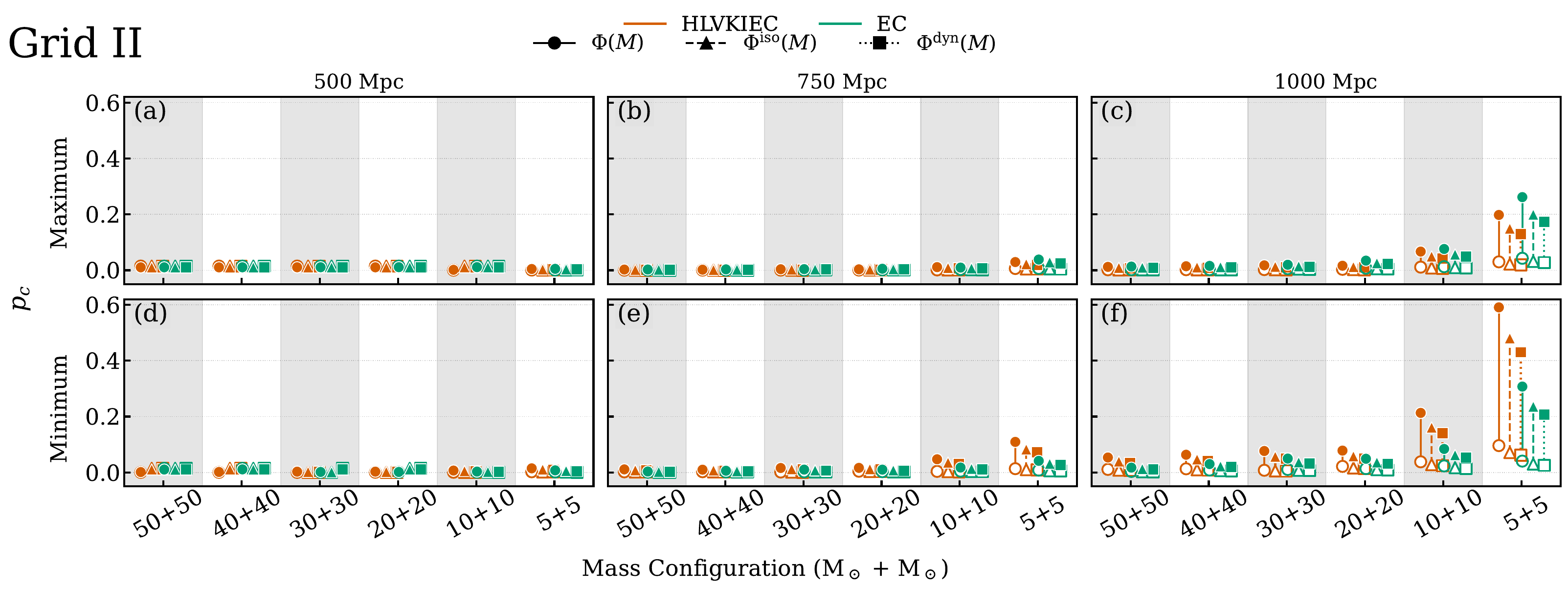}
    \caption{Probability of chance alignment, $p_c$, for Grid II BBH injections in maximum \textit{(top row, Panels a–c)} and minimum \textit{(bottom row, Panels d–f)} sensitivity sky regions, within a range of $p_c$ = 0 - 0.6. The three columns \textit{(left to right)} correspond to 500, 750, and 1000 Mpc. Colours denote the GW detector networks: HLVKIEC \textit{(red)} and EC \textit{(green)}. The line style/marker encodes the different GSMFs across BBH formation channels used in Equation~\ref{eta_eqn}: $\Phi (M)$ (\textit{solid}, Equation~\ref{schechter}), $\Phi^{\rm iso}(M)$ (\textit{dashed}, Equation~\ref{phi_iso}), $\Phi^{\rm dyn}(M)$ (\textit{dotted}, Equation~\ref{phi_dyn}). For each injection, the vertical segment spans the value at $V_{50}$ \textit{(hollow marker)} to the value at $V_{90}$ \textit{(filled marker)}. The $p_c$ values are computed assuming an injected \Mstar~host at the maximum/minimum sensitivity locations and evaluated against the NED-LVS galaxy catalogue; therefore, very low $p_c$ values (particularly at 500 Mpc) arise because no such galaxy is actually present at those sky positions.}
    \label{fig:pchance_II}
\end{figure*}

We evaluate the probability of chance alignment, $p_c$ (Section~\ref{pchance}), for the Grid II injections, shown in Figure~\ref{fig:pchance_II}. At the \emph{maximum sensitivity} locations \textit{(Panels a-c)}, both HLVKIEC and EC yield uniformly small values across all distances ($p_c < 0.01$); at 1000 Mpc, there is a slight increase for 10+10 ($p_c \approx 0.1$) and 5+5~\Msun~($p_c \gtrsim 0.2$) injections. 

At the \emph{minimum sensitivity} locations \textit{(Panels d–f)}, both networks yield small values up to 750 Mpc, with a slight increase ($p_c \sim 0.1$) for the 5+5~\Msun~injection at 750 Mpc. At 1000 Mpc, the values increase slightly ($p_c \sim 0.1$) for the heavier BBH systems and significantly for the lighter systems ($p_c \lesssim 0.2$ for 10+10~\Msun; $p_c \sim 0.2 - 0.6$ for 5+5~\Msun;)

Across panels, HLVKIEC and EC track each other closely; $p_c$ values vary very slightly with mass, sky location, and BBH formation channels. Overall, chance alignments are rare for nearby, high–mass events at favourable sky positions, but become probable for lighter binaries at larger distances and at minimum-sensitivity locations.

\section{Discussion}
\subsection{GW Detector Network Comparison} \label{networkcomparison}
\begin{figure}
    \centering
    \includegraphics[width=\linewidth]{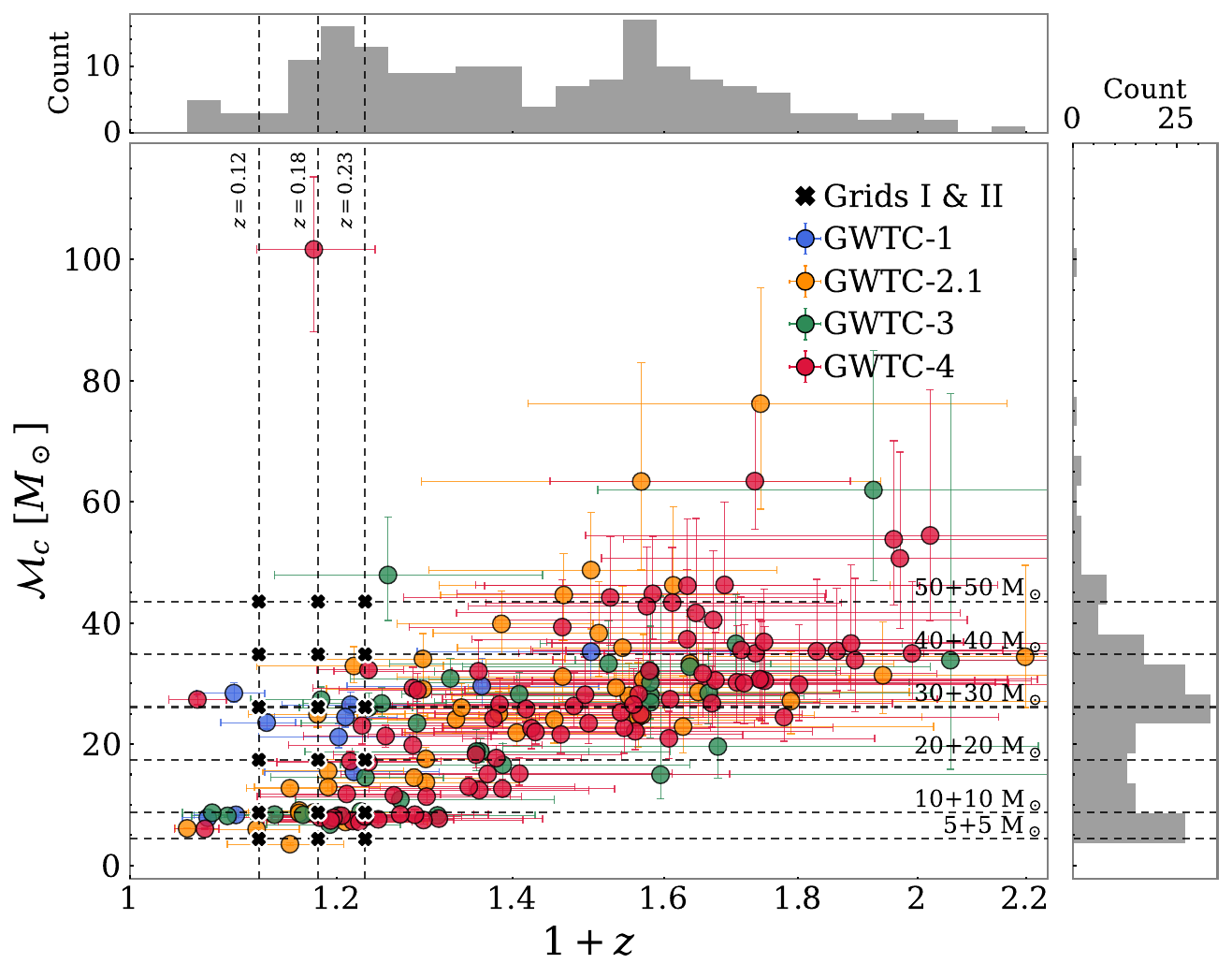}
    \caption{Source-frame chirp mass $\mathcal{M}_c$ vs $z$ for BBH events in GWTC-4.0 \citep{gwtc1, gwtc_2a, gwtc_2b, gwtc3, gwtc4} \textit{(scatter points)}. Redshifts are computed from the reported luminosity distances using the cosmology adopted in this work. Using the detector-frame chirp mass values $\mathcal{M}_{\rm c, det}$, we recompute the source-frame chirp mass as $\mathcal{M}_c = \mathcal{M}_{\rm c, det}/(1+z)$ using those redshifts. The \textit{top} and \textit{right} panels give the marginal count histograms in $z$ and $\mathcal{M}_c$. \textit{Dashed vertical lines} indicate the injection redshifts $z$ = 0.12, 0.18, and 0.23; \textit{dashed horizontal lines} indicate the equal-mass injection chirp masses for $m_1 = m_2 = (50.0, 40.0, 30.0, 20.0, 10.0, 5.0) M_\odot$. The \textit{black crosses} mark the intersections of these grids (Grids I \& II, Tables \ref{tab:injected_hosts} \& \ref{tab:bright_dark_points}), showing where our injections fall relative to the observed GW population. 
    }
    \label{fig:chirpmass_redshift}
\end{figure}

A key objective of this work is to assess the extent to which 3G GW detector networks can localize stellar-mass BBH mergers to volumes small enough to enable unique \Mstar~host identification (referred to as "host" hereafter) and, subsequently, constrain BBH formation channels and cosmological measurements such as $H_0$. Our simulated BBH merger injections extend out to 1000 Mpc ($z \sim 0.23$), sampling a region that typically overlaps with the higher SNR of the observed GW parameter space (Figure~\ref{fig:chirpmass_redshift}). Across both injection grids I and II (Section~\ref{inj_grids}) and across all distances up to 1000 Mpc, we find that incorporating 3G detectors, ET and CE, dramatically improves localization performance relative to current-generation detector networks, with localization volumes typically less than the defined theoretical comoving volume thresholds. It should be noted, however, that the comoving volume thresholds are defined as an expectation value under the assumption of Poisson statistics for the galaxy distribution; while sub-threshold localization volumes increase the probability of a unique host, they do not guarantee it. Further, we assume ET has a configuration of an equilateral triangle with 10~km arms in our simulations. \citealt{Branchesi_2023} showed that two L-shaped ET detectors with 15~km arms, separated by $\sim$1000-2000~km, is likely to provide an improved (i.e. reduced) distance uncertainty over a single triangular ET detector with 10~km arms, while additional baselines provide further improved sky-localisation. The results presented in this paper are therefore, likely conservative in the sense that this reduced uncertainty will likely lead to smaller localization volumes in the two detector configuration. To quantify this effect fully is beyond the scope of this paper.

Grid I examines how localization performance scales with the mass configurations and distances, by fixing the injections to \Mstar~galaxies at 500, 750, and 1000 Mpc (Table~\ref{tab:injected_hosts}). Within this framework, both HLVKIEC and EC networks consistently achieve sub-threshold localization volumes (Figure~\ref{fig:vol_comp_I}). In contrast, the HLV network produces substantially larger localization volumes, with $V_{90}^{\rm HLV} \sim 10^2 - 10^5~\rm{Mpc}^3$, and thus rarely supports confident host associations or formation channel-based constraints. These results must be interpreted in the context of the underlying FIM-based PE method that is valid in the high-SNR limit, where posterior distributions are sharply peaked and approximately Gaussian. These conditions are met by all injections in the HLVKIEC and EC networks, but not for all the HLV injections. 



In addition to Grid I, Grid II explores the role of detector geometry by sampling sky locations within the most and least sensitive regions of the EC and HLVKIEC antenna patterns (Figure~\ref{fig:antennapattern}). This enables a direct assessment of how localization volumes vary with directional sensitivity (Figure~\ref{fig:vol_comp_II}). Both HLVKIEC and EC networks exhibit degraded localization in the "dark" regions of their antenna patterns. Importantly, the comparison between the networks depends on how these minimum- and maximum-sensitivity sky positions are defined. These locations are computed independently for each network and as a result, the minimum-sensitivity sky position of HLVKIEC does not necessarily correspond with the combined minimum of ET and CE individually. In such cases, HLVKIEC may appear to perform worse than EC (\textit{Panels d - f} in Figure~\ref{fig:vol_comp_II}), but this is a consequence of the choice of these sky positions.

Together, the Grid I and Grid II results show that future GW detector networks will enable targeted galaxy follow-up campaigns for a significant fraction of BBH mergers. In the best cases, especially for massive BBH mergers, the localization volumes are small enough to isolate a unique host. This, in turn, opens a pathway for associating BBH mergers with specific galaxy populations and statistically distinguishing between formation scenarios, such as isolated evolution and dynamical assembly.



\subsection{Implications for Host Identification and Constraining BBH Formation Channels} \label{implications}

\subsubsection{BBH Formation Channel Constraints from Theoretical Comoving Volume Thresholds}
In addition to assessing the feasibility of host identification, comparisons between the simulated localization volumes and the theoretical thresholds $V_{\min}$ (Equation~\ref{vmin_eqn}), $V^{\rm iso}_{\min}$, and $V^{\rm dyn}_{\min}$ provide insights into the underlying BBH formation channels (Figures~\ref{fig:vol_comp_I} and \ref{fig:vol_comp_II}). As defined in Section~\ref{theoreticalvol}, each threshold volume represents the minimum comoving volume required to contain, on average, one \Mstar~galaxy, assuming a specific BBH formation channel. $V_{\min}$ corresponds to the unweighted GSMF and serves as a general baseline, while $V^{\rm iso}_{\min}$ and $V^{\rm dyn}_{\min}$ incorporate mass-weighted GSMFs for the isolated and dynamical channels, respectively.

When the credible localization volumes, $V_{50}$ and $V_{90}$, for a given BBH injection lie below a threshold, host identification may be achieved on an event-by-event basis (under the single \Mstar~host assumption). In contrast, formation-channel inference requires population-level evidence. For instance, a localization volume that falls below $V^{\rm dyn}_{\min}$ but exceeds $V^{\rm iso}_{\min}$ is, on average, more consistent with the dynamical channel than the isolated one. This does not imply definitive channel classification on an event-by-event basis, but rather indicates which channels are statistically compatible with the inferred host environment, assuming a single \Mstar~host. It is important to note that the dynamical channel considered here refers specifically to GCs. As such, it is plausible that a given galaxy could support BBH formation through both isolated and dynamical pathways. 

\subsubsection{Metallicity-Based Constraints}
An additional layer of constraint comes from the metallicity dependence of BBH formation. Isolated binary evolution models predict a strong bias toward low-metallicity environments \citep[e.g.][]{giacobbo2018, Spera_2019}, while dynamical formation channels are comparatively insensitive to progenitor metallicity \citep[e.g.][]{2022mapelli}. By adopting a representative metallicity threshold of $12 + \rm{log(O/H)} = 8.3$ (Section~\ref{metallicity}), we derive a corresponding minimum localization volume, $V^Z_{\min} = 54.42~\rm{Mpc}^3$, below which we expect to detect, on average, one such low-metallicity galaxy. Almost all of the BBH injections localized by the HLVKIEC and EC networks meet this criterion, particularly for high-mass BBH mergers at nearby distances. At higher distances, the low-mass BBH mergers have localization volumes $\gtrsim V^Z_{\min}$ in both Grids I and II. We note that in the low metallicity scenario it may not be possible to identify the host galaxy directly, because there may be fainter, even lower metallicity hosts. However, the absence of bright galaxies in such small localisations would rule out the mergers occurring in such hosts. 

\subsubsection{Mass Fractions and the Role of Galaxy Catalogue Completeness} \label{massfraction_galcat_incomp}
We further assess the prominence of the injected host within the GW localization volume using two complementary mass fractions: $\mathbb{M}_{\rm *, \Phi}$ and $\mathbb{M}_{\rm *, obs}$ (Section~\ref{theoretical_mass}, Figures~\ref{fig:gridI_massratio} and \ref{fig:gridII_massratio}). The former quantifies the ratio of the mass of the host to the total stellar mass theoretically expected from the GSMF within the localization volume, while the latter compares the host mass to the total observed stellar mass of galaxies crossmatched with the NED-LVS galaxy catalogue. A value greater than one for either mass fraction increases the plausibility of host identification. In the context of this work, where the injected galaxies are known a priori, it indicates successful recovery of the injected \Mstar~host. 

Although $\mathbb{M}_{\rm *, obs}$ is defined to quantify the observed stellar-mass content within the localization volume, it is governed entirely by the completeness of the underlying galaxy catalogue. For nearly all the injections, we calculate $\mathbb{M}_{\rm *, obs} \sim 1$, indicating that the crossmatch recovers only the injected host. In contrast, the theoretical mass fraction values, $\mathbb{M}_{\rm *, \Phi}$, are systematically larger, implying that the total stellar mass expected within the localization volume is significantly under-represented in the crossmatched sample. This reflects the incompleteness of the NED-LVS galaxy catalogue at these distances. Further, the incompleteness is also non-uniform: for instance, at 750~Mpc, the 5+5~\Msun~injection yields $\mathbb{M}_{\rm *, obs} \lesssim$ 0.1, whereas at 1000~Mpc, the same mass configuration has a value of $\mathbb{M}_{\rm *, obs}$ = 1. This distance-dependent, non-uniform recovery of galaxies demonstrates that $\mathbb{M}_{\rm *, obs}$ cannot be meaningfully interpreted without modelling the galaxy catalogue completeness and catalogue selection function.


\subsubsection{Constraints from Chance Alignment Probability}
Next, we incorporate the probability of chance alignment, $p_c$ (Section~\ref{pchance}), to assess whether a galaxy consistent with a given BBH formation channel could arise by random association. As shown in Figures~\ref{fig:pchance_I} and \ref{fig:pchance_II}, $p_c$ values computed under $\Phi^{\rm iso}(M)$ and $\Phi^{\rm dyn}(M)$ follow a consistent ordering, $p_c^{\rm dyn} \lesssim p_c^{\rm iso} \lesssim p_c$, though the differences between channels are typically modest. $p_c$ serves as a valuable consistency check: low values ($\lesssim 0.1-0.2$) indicate that a candidate host is unlikely to be a chance superposition, thereby increasing confidence in its association with the BBH merger. Conversely, high $p_c$ values across all GSMFs, as seen for HLV at large distances, indicate that no confident host association and no channel-based inference is possible. While $p_c$ does not independently distinguish formation channels, it improves the reliability of host associations when combined with the theoretical volume thresholds, metallicity cuts, and mass fractions. 

\subsubsection{Event Rates of BBH Mergers with Identifiable Hosts}
Further, we can estimate an event rate per year of BBH mergers with identifiable hosts out to $z \sim 0.23$ in the 3G era. The current generation of GW detectors (HLVK) detects a subset of the entire BBH population with SNR $\geq$12 at $z \lesssim 0.1$, with detection efficiency depending on the BH masses and the orientation of the binary with respect to our line-of-sight. In comparison, \citealt{iacovelli2022} estimate that ET alone can detect 100\% of the BBH mergers with SNR $\geq$12 up to $z=1$. From our results, we find that up to $z = 0.23$, unique host associations are possible, i.e., $f_{\rm host} \approx 1$. Together, this implies that the entire population of BBH mergers and their associated hosts are detectable and potentially identifiable up to $z = 0.23$ in the 3G era. The event rate of BBH mergers with identifiable unique hosts per year is then
\begin{equation}
    \displaystyle \mathcal{R}_{\rm host} = \mathcal{R}_{\rm BBH}(z)~V_c(z)~\mathcal{D}~f_{\rm host},
\end{equation}
where $\mathcal{R}_{\rm BBH}$ is the volumetric BBH merger rate. BBH merger rates are highly uncertain (see \citealt{2022mandelbroek} and \citealt{2023sedda} for a review). LVK inferred the local ($z \approx 0$) BBH merger rate to be 14 - 26 $\rm Gpc^{-3}~yr^{-1}$ \citep{gwtc4}. Isolated BBH formation channels suggest a volumetric rate of $\mathcal{R}_{\rm BBH} \sim 0.5 - 5 \times 10^3~\rm{Gpc^{-3} yr^{-1}}$ at $z \lesssim 1$, while dynamical formation channels suggest $\mathcal{R}_{\rm BBH} \sim 10^{-3} -10^2~\rm{Gpc^{-3} yr^{-1}}$ at similar redshifts \citep{2022mandelbroek, 2023sedda}. Here, we assume $\mathcal{R}_{\rm BBH} \sim 50~\rm{Gpc^{-3} yr^{-1}}$ at $z \lesssim 0.23$. $V_c$ is the comoving volume, and at $z = 0.23$, $V_c = 3.45~\rm{Gpc^{3}}$. $\mathcal{D}$ = 0.85 is the assumed duty cycle for ET in 1 year \citep{etbluebook}. Under these assumptions, the expected number of BBH mergers with SNRs $\geq$12 with a uniquely identifiable host detected by ET alone is $\mathcal{R}_{\rm host} \sim 100~\rm {yr^{-1}}$ at $z \lesssim 0.23$, consistent with estimates presented in figure 18 of \citet{Branchesi_2023} and figure 2 of \citet{2024ish}.



\subsubsection{Dependence on Host Galaxy Mass}
We emphasize that this work was done under the assumption of a BBH merger occurring in a \Mstar~galaxy. If instead we simulated the BBH mergers in lower mass galaxies (e.g. $10^8$~\Msun), the corresponding $V_{\rm min}$ values would be $\approx 10^2~\rm{Mpc^3}$ for the distances considered here (Figure~\ref{fig:vmin_comparison}). The localization volumes achieved by the HLVKIEC and EC detector networks remain well below this threshold (Figures~\ref{fig:vol_comp_I} and \ref{fig:vol_comp_II}), indicating that unique host identification would still be feasible, even for dwarf-galaxy hosts. 

\subsection{Special Cases}
In addition to the bulk of the BBH population which we might expect to form through standard binary evolution channels and hence to arise in otherwise normal galaxies, there are additional special cases that may arise in which the properties of the environment may be sufficiently unusual that they can be identified, even in larger uncertainty regions. These are the formation of BBH systems in AGN accretion discs, and cases in which the GW emission itself is lensed. A full consideration of these scenarios is beyond the scope of this paper, in principle the volumetric densities of both AGN and lensing systems are much lower than that of galaxies, such that such identifications should be more straightforward. We briefly consider each in turn. 

In the case of BBH mergers formed through gas capture in AGN accretion discs \citep[e.g.][]{mckernan20}, we would expect to observe a clear AGN within the volumetric error boxes. Even in large current generation error boxes some claims have been made for AGN association based outbursts observed in the months after a BBH merger \citep{agn_bbh_det}, although such claims are controversial. However, as error boxes get smaller, the probability of observing an AGN goes down. Even with 2G detectors it should be possible to identify AGN in some cases \citep{bartos17}. The minimum volume estimates for a AGN scenario in the 3G case will depend on assumptions about BBH formation rate as a function of AGN luminosity. However, even for faint AGN they are likely to be larger $\sim 1000$ Mpc$^3$, suggesting that such associations will be straightforward for both EC and HLVKIEC configurations. 

Similarly, the identification of lensed GW signals of significant interest for a raft of scientific questions \citep[see e.g.,][]{smith25}. Ideally, such a lensing detection would consist of multiple detections of the same GW signal with variable magnification. However, Malmquist-bias may mean that it is sometimes only the most highly magnified events that are detected. The time delay between signals depends on the lensing mass and geometry. Large galaxy or cluster size lensing should be readily apparent in error boxes from EC or HLVKIEC configurations. Smaller scale lensing (shorter time delays) may not be, especially in cases where the angular separation of the multiple images is not resolved in existing imaging (i.e. where one cannot visually identify the lens). Nonetheless, the identification of lensing systems based on EM observations only, with no multiple GW detections at the time, may provide a route to enhancing the lensing return, in particular if these observations provide a route of searching for additional lensed signatures at lower SNRs.

\section{Conclusion} \label{conclusion}
As of March 2025, the LVK collaboration has reported $\sim 210$ detections of BBH mergers ($\rm p_{astro} \geq 0.5$; \citealt{gwtc4}), since the first GW detection of a BBH merger in 2015. The lack of EM emission from BBH mergers, combined with the large GW localization areas produced by the current-generation detector networks, makes the identification of a host to a BBH merger difficult. A confident host association would have several important astrophysical and cosmological implications, such as constraining BBH formation channels and providing an independent $H_0$ estimate. The planned addition of LIGO-India \citep{indigo1, indigo2, 2024ligoa} and 3G detectors such as ET \citep{Punturo_2010, 2011hild, 2020maggiore, etbluebook} and CE \citep{2019reitze, 2021evans, 2023evans} is expected to reduce GW sky localization areas and improve distance precision for nearby, high-SNR events. 

In this work, we simulate BBH mergers injected into \Mstar~galaxies to assess the feasibility of such a host identification with future GW detector networks. We construct two complementary grids of injections: Grid I explores how localization performance scales with source properties such as mass and distance, while Grid II incorporates directional sensitivity through antenna patterns to evaluate sky-position dependence. We evaluate the inferred GW localization volumes from each injection using the FIM formalism as implemented within the \texttt{BILBY} framework, for three detector networks - HLV, HLVKIEC, and EC - and compare them to theoretical comoving volume thresholds derived from the GSMF, including thresholds computed using GSMFs weighted by assumptions about BBH formation via isolated binary evolution and dynamical assembly in dense stellar environments such as GCs. 

To assess the implications for host identification and constraining the BBH formation channels, we introduce several diagnostics: (i) statistical volume thresholds ($V_{\min}$, $V^{\rm iso}_{\min}$, $V^{\rm dyn}_{\min}$) corresponding to different BBH formation channels, (ii) a representative metallicity threshold motivated by progenitor environment constraints ($V^Z_{\min}$), (iii) the prominence of the injected host within the localization volume via mass fractions ($\mathbb{M}_{*, \Phi}$ and $\mathbb{M}_{\rm *, obs}$), and (iv) the probability of chance alignment ($p_c$). 

We find that 3G detector networks such as EC and HLVKIEC enable significant improvement in localization, achieving localisation volumes smaller than the theoretical thresholds for distances up to $\sim$ 1000 Mpc. Under these conditions, unique host associations become feasible, at a rate of $\sim 100~{\rm yr^{-1}}$. Additionally, sub-threshold volumes enable statistical differentiation between isolated and dynamical BBH formation channels in favourable cases. 

Future extensions of this work could include expanding the injection grid to consider galaxies of different stellar masses, out to higher redshifts ($z \sim 2$), and sampling a broader range of GW source parameters, such as mass fraction, spin, inclination, and eccentricity (Appendix~\ref{gwparams}); incorporating models of galaxy catalogue completeness and selection effects; and applying the proposed metrics to real GW detections to evaluate candidate host associations. Additional BBH formation channels such as chemically homogeneous evolution, mergers in the accretion disks of AGNs, and hierarchical mergers, as well as mixed-channel scenarios involving multiple pathways within a single HG, also merit further investigation.

\section*{Acknowledgements}
We are thankful to Christopher Berry for his useful comments on our draft. SB acknowledges studentship support from the Dutch Research Council (NWO) under the project number 680.92.18.02. PGJ is supported by the European Union (ERC, StarStruck, 101095973). Views and opinions expressed are however those of the author(s) only and do not necessarily reflect those of the European Union or the European Research Council. Neither the European Union nor the granting authority can be held responsible for them. This work made use of Python packages \texttt{NUMPY} \citep{numpy}, \texttt{SCIPY} \citep{scipy}, and \texttt{MATPLOTLIB} \citep{matplotlib}. This work made use of \texttt{ASTROPY}: a community-developed core Python package and an ecosystem of tools and resources for astronomy \citep{astropy:2013, astropy:2018, astropy:2022}. The authors are grateful for computational resources provided by the LIGO Laboratory and supported by National Science Foundation Grants PHY-0757058 and PHY-0823459. 
\section*{Data Availability}
Relevant data and code will be made available in a reproduction package uploaded to Github at the following URL: \hyperlink{https://github.com/sumedhabiswas/popnet}{https://github.com/sumedhabiswas/popnet}.



\bibliographystyle{mnras}
\bibliography{example} 




\appendix

\section{Dependence of GW Localization on Source Parameters and Consistency with GW Observations} \label{gwparams}
\begin{figure}
    \centering
    \includegraphics[width=\linewidth]{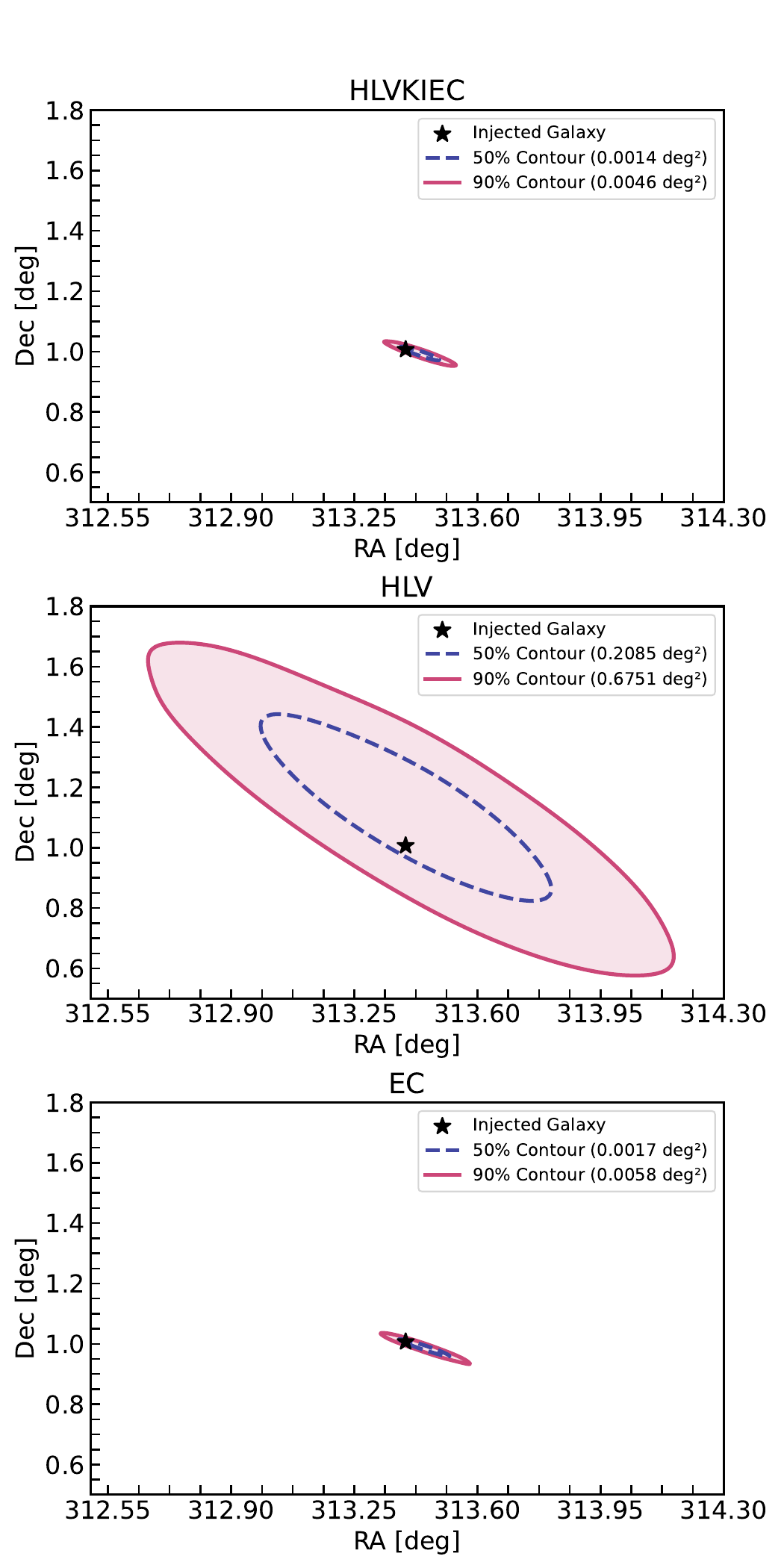}
    \caption{Skymaps for the 50+50~\Msun~injection at 1000 Mpc (RA = 313.4$^{\circ}$, Dec = 1.0$^{\circ}$; \textit{black star}), as recovered by the 3 detector networks: HLVKIEC \textit{(top)}, HLV \textit{(middle)}, and EC \textit{(bottom)}. Each panel shows the 50\% \textit{(dashed indigo)} and 90\% \textit{(solid magenta)} credible regions.}
    \label{fig:skymap}
\end{figure}

In our \texttt{BILBY} simulations, BBH mergers are modelled using fixed GW source parameters, as summarized in Tables~\ref{tab:injection_parameters} and \ref{tab:prior_ranges}. This subsection evaluates how these modelling assumptions influence our analysis outcomes and compares the chosen parameter values to those inferred from observed GW events. As an example, Figure~\ref{fig:skymap} shows the 2D localization areas of a 50+50~\Msun~Grid I injection at 1000 Mpc. 

\subsection{Eccentricity}
Eccentricity is an important parameter, both for distinguishing BBH formation channels \citep[e.g.][]{2016nishi, 2017nishi, 2016breivik, 2021zevin_e_bbhform} and understanding its impact on GW localization regions \citep[e.g.][]{baosan_2015, 2017ma_ecc_loc, 2024yang}. Dynamical channels, such as those involving GCs or hierarchical triples, can retain measurable eccentricity by the time the binary enters the ground-based GW detector band, especially at lower GW frequencies \citep[e.g.][]{amico2024}. In contrast, BBHs formed through isolated binary evolution are expected to circularize efficiently via gravitational radiation, making the assumption of quasi-circular orbits well justified in that scenario \citep{peters1964}. In our simulations, we use the \texttt{IMRPhenomXPHM} waveform approximant \citep{Pratten_2021}, which assumes quasi-circular (i.e., non-eccentric) binaries. While this is appropriate for the isolated channel, 
previous studies have shown that even moderate eccentricity can improve localization performance. \cite{2017ma_ecc_loc} find that increasing the eccentricity from 0 to 0.4 improves localization by a factor of $\sim$2 for a binary with a total mass of 100~\Msun, and by a factor of $\sim$1.3 for a GW150914-like system (total mass of 65~\Msun; \citealt{gwtc_2b}). For lower-mass binaries such as GW151226 (total mass of 22~\Msun; \citealt{gwtc_2b}), the improvement is negligible. \cite{Gondan_2018} report similar trends: for a 30+30~\Msun~non-spinning BBH with high eccentricity (pericenter distance of 20~$M_{\rm tot}$), localization improves by a factor of $\sim$2. \cite{pan2019} also demonstrated that the accuracy of sky localization increases appreciably with eccentricity for binaries with total mass $M \geq 40~M_\odot$, while the effect is negligible for lower-mass systems. These findings suggest that our localization estimates for dynamically formed BBHs, particularly at higher masses, may represent conservative lower limits, as the inclusion of eccentricity would likely enhance localization performance beyond what is captured in our quasi-circular simulations. Consequently, for dynamically formed BBHs with non-negligible eccentricity, incorporating eccentric waveforms in future analyses may not only improve localization accuracy but also increase the likelihood of successful host identification.


\subsection{Mass Ratio ($q$)}
In our injections, we adopt equal-mass binaries with primary and secondary BH masses drawn from the set $[50.0, 40.0, 30.0, 20.0, 10.0, 5.0]~M_\odot$, resulting in a fixed mass ratio of $q = 1$ across all injections. This choice is observationally motivated: analyses of the GWTC-3.0 catalogue \citep{2023_gwtc3_pop, gwtc3} suggest that BBH systems with symmetric mass ratios $q \gtrsim 0.7$ dominate the population, particularly among high-mass mergers \citep{tomoya}. 
As such, our adopted mass ratio represents a physically plausible configuration within the observed BBH population. For further context, in Figure~\ref{fig:chirpmass_redshift}, we calculate and plot the source-frame chirp masses\footnote{The error bars on chirp mass are accurate to $\sim$0.1-1\% \citep{1994cutler}, dependent on the waveform model, and degenerate with spin and the $\mathcal{M}_c - z$ degeneracy \citep{2014messenger}.} $\mathcal{M}_c = (m_1 m_2)^{3/5}/(m_1 + m_2)^{1/5}$ of our equal-mass injections as a function of redshift, and compare them to GW observations (GWTC-4.0; \citealt{gwtc1, gwtc_2a, gwtc_2b, gwtc3, gwtc4}). 

\subsection{BH Spins ($a_1, a_2$)}
In our injections, we fix the dimensionless spin magnitudes of both BHs to $a_1 = a_2 = 0.1$, corresponding to nearly non-spinning BHs. This is motivated by population-level analyses of the GWTC-3.0 catalog \citep{gwtc3}, which suggest that low spin magnitudes are preferentially associated with symmetric mass-ratio binaries ($q = 1$) \citep{2023_gwtc3_pop} (see their figure~21). Our adopted spin and mass ratio values are therefore broadly consistent with current observational constraints. 

\subsection{Binary Inclination Angle ($\theta_{\rm JN}$)}
In all our injections, we fix the binary inclination angle to $\theta_{\rm JN} = 0.4$ radians, corresponding to a nearly face-on orientation. Such configurations yield GW signals with higher SNRs and tighter posterior distributions, leading to reduced localization volumes. Therefore, our choice of $\theta_{\rm JN} = 0.4$ radians approximates a best-case scenario for source localization.




\bsp	
\label{lastpage}
\end{document}